\documentclass[sigplan,screen]{acmart}
\usepackage{tikz}
\usepackage{amsmath}
\usepackage{printlen}
\usepackage{blindtext}
\usepackage{subcaption}
\usepackage{tabularx}
\usepackage{algorithm}
\usepackage{algpseudocode}
\usepackage{pifont}
\usepackage{balance}
\usepackage[]{hyperref}

\providecommand{\sysname}{\textsc{Aqua}\xspace}

\providecommand{\sysnameT}{\textsc{\sysname Tensors}\xspace}
\providecommand{\sysnamePl}{\textsc{\sysname-placer}\xspace}
\providecommand{\sysnameprofiler}{\textsc{\sysname-profiler}\xspace}
\providecommand{\sysnamelib}{\textsc{\sysname-lib}\xspace}

\providecommand{\nvlinks}{\textsc{Nvlinks}\xspace}

\providecommand{\vs}{vs. }
\providecommand{\ie}{\emph{i.e.,} }
\providecommand{\eg}{\emph{e.g.,} }

\providecommand{\etc}{\emph{etc{}}}      

\providecommand{\myparab}[1]{\vspace{1pt}\noindent\textbf{#1} }

\usepackage[small,compact]{titlesec}
\titlespacing\section{0pt}{4pt plus 2pt minus 2pt}{2pt plus 2pt minus 2pt}
\titlespacing\subsection{0pt}{2pt plus 2pt minus 2pt}{2pt plus 2pt minus 2pt}

\AtBeginDocument{%
}

\copyrightyear{2025}
\acmYear{2025}
\setcopyright{rightsretained}
\acmConference[ASPLOS '25]{Proceedings of the 30th ACM International Conference on Architectural Support for Programming Languages and Operating Systems, Volume 2}{March 30-April 3, 2025}{Rotterdam, Netherlands}
\acmBooktitle{Proceedings of the 30th ACM International Conference on Architectural Support for Programming Languages and Operating Systems, Volume 2 (ASPLOS '25), March 30-April 3, 2025, Rotterdam, Netherlands}\acmDOI{10.1145/3676641.3715983}
\acmISBN{979-8-4007-1079-7/2025/03}

\makeatletter
\gdef\@copyrightpermission{
  \begin{minipage}{0.2\columnwidth}
   \href{https://creativecommons.org/licenses/by/4.0/}{\includegraphics[width=0.90\textwidth]{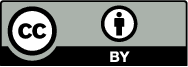}}
  \end{minipage}\hfill
  \begin{minipage}{0.8\columnwidth}
   \href{https://creativecommons.org/licenses/by/4.0/}{This work is licensed under a Creative Commons Attribution International 4.0 License.}
  \end{minipage}
  \vspace{5pt}
}
\makeatother

\begin{document}

\ccsdesc[500]{Computing methodologies~Machine learning}
\ccsdesc[500]{Networks~Data center networks}
\ccsdesc[500]{Networks~Data path algorithms}

\keywords{Large language models; Generative AI; Paging; Virtual memory; GPU interconnects; Memory offloading;}

\title{\sysname: Network-Accelerated Memory Offloading for LLMs in Scale-Up GPU Domains}

\author{Abhishek Vijaya Kumar}
\affiliation{
  \institution{Cornell University}
  \city{Ithaca}
  \country{USA}
}
\email{abhishek@cs.cornell.edu}

\author{Gianni Antichi}
\affiliation{
  \institution{Politecnico di Milano}
  \city{Milan}
  \country{Italy}
}
\email{gianni@polimi.it}

\author{Rachee Singh}
\affiliation{
  \institution{Cornell University}
  \city{Ithaca}
  \country{USA}
}
\email{rachee@cs.cornell.edu}

\begin{abstract}
Inference on large-language models (LLMs) is constrained by GPU memory capacity. A sudden increase in the number of inference requests to a cloud-hosted LLM can deplete GPU memory, leading to contention between multiple prompts for limited resources. Modern LLM serving engines deal with the challenge of limited GPU memory using admission control, which causes them to be unresponsive during request bursts. We propose that preemptive scheduling of prompts in time slices is essential for ensuring responsive LLM inference, especially under conditions of high load and limited GPU memory. However, preempting prompt inference incurs a high paging overhead, which reduces inference throughput. We present \sysname, a GPU memory management framework that significantly reduces the overhead of paging inference state; achieving both responsive and high throughput inference even under bursty request patterns. We evaluate \sysname by hosting several state-of-the-art large generative ML models of different modalities on servers with 8 Nvidia H100 80G GPUs. \sysname improves the responsiveness of LLM inference by $20\times$ compared to the state-of-the-art. It improves LLM inference throughput over a single long prompt by $4\times$.\footnote{\sysname's code is available
at \url{https://github.com/aquaml}.} 

\end{abstract}

\maketitle 

\section{Introduction}

%

Recent years have seen an explosive growth in popularity of generative machine learning (ML) models for a multitude of tasks ranging from text summarization to multimedia content creation~\cite{openai_customer_stories,azure_openai_blog, 
aws_generative_ai_use_cases,mewtant_2024}. These models are typically hosted on virtualized multi-GPU cloud servers~\cite{amazon-sagemaker,azure-ai-studio, together_ai,openai-api}, which interconnect a handful of ML accelerators on the board of the server with high-bandwidth links~\cite{nvlink}. 

\myparab{Inference is memory intensive.}
Generative ML models have a massive memory footprint~\cite{codellama,
mistral,mixtral,mpt}. For example, hosting the Llama-405B model consumes 
800 gigabytes of GPU high-bandwidth memory (HBM). Moreover, large-language models
(LLMs) generate a key-value (KV) cache while processing input prompts
using the attention mechanism~\cite{attention}. The KV cache 
grows quadratically with the length of an input prompt and its generated sequence, 
taking several gigabytes of GPU memory per prompt~\cite{vllm}. 
Thus, a sudden increase in the number of inference requests to the LLM can 
deplete the GPU memory, leading to memory contention as multiple prompts 
compete for GPU memory.

\begin{figure}[h]
    \centering
    \begin{subfigure}[b]{0.23\textwidth}
    \includegraphics[width=\columnwidth]{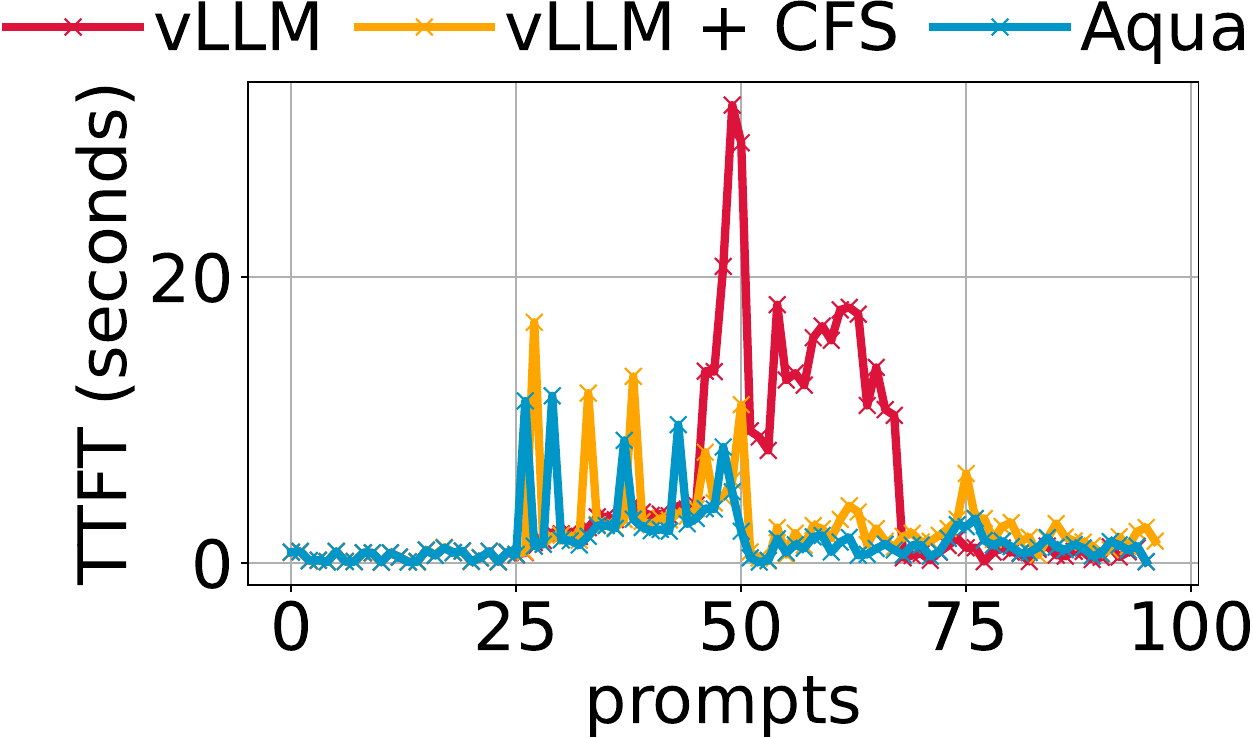}
    \caption{\small TTFTs}
    \label{fig:cfs_mb_ttft}
    \end{subfigure}
    \begin{subfigure}[b]{0.23\textwidth}
    \centering
    \includegraphics[width=\columnwidth]{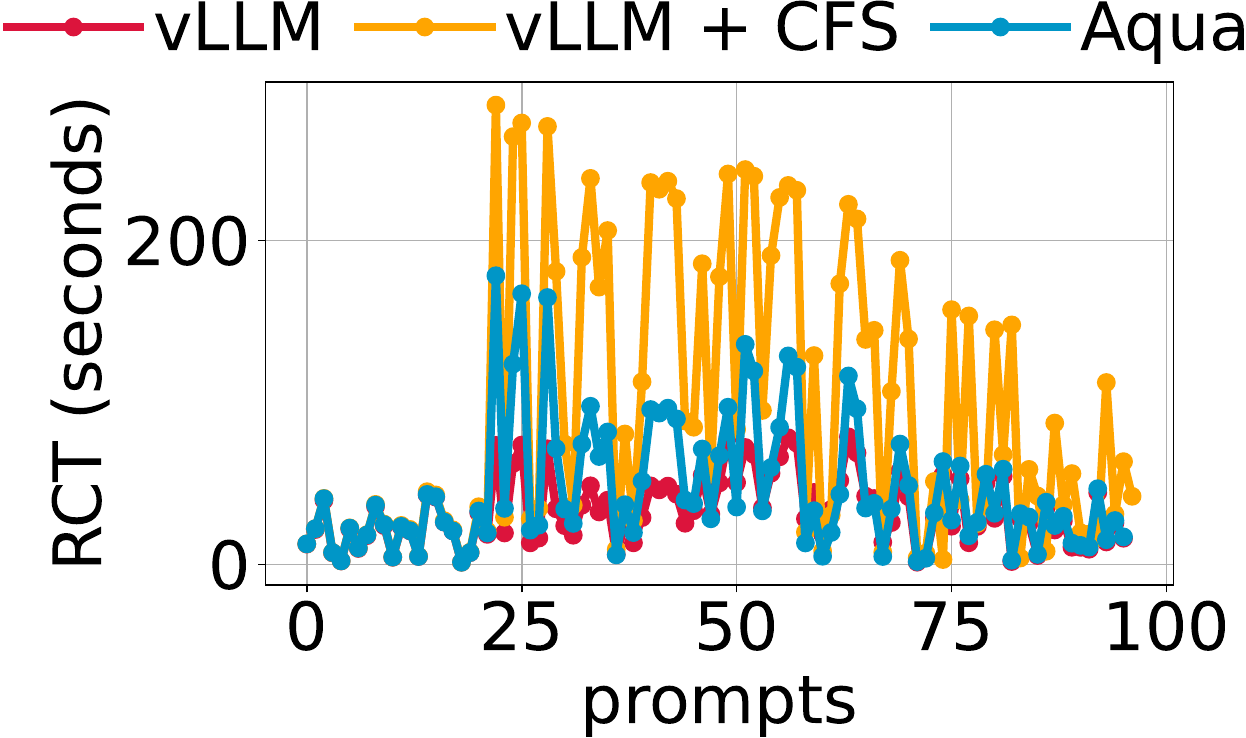}
    \caption{\small RCT}
    \label{fig:cfs_mb_rct}
    \end{subfigure}
    \vspace{-1em}
    \caption{\small Responsiveness (measured using time-to-first-token or TTFT) and 
    throughput (measured using request completion time or RCT) of inference queries on LLMs. 
    Since vLLM batch processes queries, it has low RCT (high throughput) 
    but high TTFT (low responsiveness). Fair-scheduling queries improves 
    responsiveness but paging overheads dominate RCT. \sysname reduces 
    paging overheads by offloading memory over high-speed multi-GPU 
    interconnects (\eg \nvlinks), achieving responsive inference with 
    low RCT.}
    \vspace{-1em}
    \label{fig:cfs_mb}
\end{figure}

\myparab{Managing bursts in inference requests with queueing.}
Modern LLM serving engines deal with the challenge
of limited GPU memory using admission control~\cite{vtc, vllm}.
They schedule a batch of prompts for inference and queue 
the remaining until free GPU memory is available.
However, this paradigm of batch processing with admission control, 
is susceptible to head-of-line blocking~\cite{hol} --- a situation where processing 
a batch of prompts delays responses to other prompts that await their turn 
in the queue.

\myparab{Queueing leads to unresponsiveness.}
We experimentally demonstrate that queuing prompts can severely degrade the responsiveness of an LLM. We serve an LLM on an Nvidia A100 GPU with 
80GB HBM using the popular serving engine 
vLLM~\cite{vllm}. We initially issue 25 prompts to vLLM, then double the request rate for one minute. This surge of requests exhausts vLLM's KV cache capacity, resulting in subsequent prompts being queued. This causes \emph{time-to-first-token} (TTFT), a measure of how quickly 
the model starts responding, to spike by 4X (red line in Fig.~\ref{fig:cfs_mb_ttft}). A different approach to handling bursts is 
to autoscale the model across additional GPUs~\cite{llumnix}. 
However, scaling out can take several minutes due to virtual 
machine startup delays, during which the model will be 
unresponsive. Since even 3 seconds of unresponsiveness causes users to abandon services~\cite{google-user-ditch}, delays of tens of seconds increase the risk of losing users.

\myparab{Our proposal: preemptable inference.}
In this work, we posit that preemptive scheduling of 
prompts in time slices is essential for achieving responsive LLM inference,
particularly under high load and limited GPU memory.
Under this scheduling scheme, each prompt will receive time slices of 
GPU compute, allowing the model to stay responsive. A preemptive scheduler 
will stop execution of the current prompt at the time slice boundary 
and page out its relevant state from GPU memory to make space for 
executing the next prompt. Preemptive scheduling with time slices 
will also unlock the door to implementing fine-grained fair scheduling 
in existing inference serving systems, which only support
coarse-grained fairness through admission control~\cite{vtc}.

\myparab{Preemption is slow due to paging overheads.}
However, preempting prompt inference requires paging the prompt's 
state out of GPU memory at the boundary of time slices.
If the inference state of a prompt is paged to host DRAM~\cite{vllm,flexgen}, 
transferring it back to GPU memory incurs a high overhead. 
The overhead of transferring state between DRAM and GPU memory 
is lower-bounded by the bandwidth of Peripheral Component 
Interconnect Express (PCIe) connections. In fact, 
we implement a preemptive fair scheduler in vLLM and find that while preemptive
scheduling improves the time-to-first-token in vLLM (orange line 
Figure~\ref{fig:cfs_mb_ttft}), it causes request completion 
time or the time to finish inferring a prompt, to 
increase by more than $2X$ (orange line in Figure~\ref{fig:cfs_mb_rct}). 

\myparab{\sysname: preemptable inference with low overhead.}
In this work, we build \sysname, a memory management framework for 
GPUs, that significantly reduces the overhead of paging 
inference state in and out of GPU memory to achieve efficient preemptive 
prompt scheduling. Reduced paging overheads enable \sysname to 
get the best of both worlds --- responsive inference (Figure~\ref{fig:cfs_mb_ttft}) 
for individual prompts while still achieving high inference 
throughput (Figure~\ref{fig:cfs_mb_rct}).

\myparab{\sysname's key insight.} 
Inference on ML models can be limited by memory capacity, memory 
bandwidth or compute, depending on the model type (\eg vision \vs text) 
and the inference query load~\cite{flexgen,hetegen2024,llumnix}. As a 
result, at any time, some inference jobs in a cloud cluster 
underutilize GPU memory, while others face memory bottlenecks.
Using this insight, \sysname decouples memory allocation from 
compute allocation by enabling GPUs to share spare memory with 
those that need more. \sysname limits these memory offloads to GPUs within a \emph{scale-up domain} in cloud datacenters. GPUs in a scale-up domain are connected via high-bandwidth \nvlinks, allowing \sysname to achieve faster access to spare memory than traditional host DRAM over PCIe, significantly reducing preemption and paging overheads. In modern datacenters, multi-accelerator servers define a scale-up domain. Recently, the scale-up domain has expanded to encompass entire racks, like in Nvidia Blackwell’s 72-GPU architecture, where high-bandwidth \nvlinks interconnect all 72 GPUs. These advancements extend the benefits of \sysname's decoupled memory allocation to entire racks~\cite{nvidia_blackwell_2024}.

\myparab{Technical contributions.}
\sysname has three components: \sysnameprofiler profiles
the characteristics of incoming inference load and GPU memory 
utilization of hosted ML models in a deployment. Using this profile,
\sysnameprofiler labels models 
as producers or consumers and feeds
this information to \sysnamePl. Producers do not fully 
utilize GPU memory and can offer the excess to other models.
Consumers are bottlenecked on GPU memory and need more.
\sysnamePl places models such that producers are paired 
with appropriate consumers within the same \nvlinks-domain. 
Finally, \sysnamelib is a library that dynamically
manages tensors offloaded from consumers to
producers' GPU memory. 

\myparab{Preemptive fair scheduling with low overheads.} 
Using \sysname to manage memory between GPUs, we build a 
preemptive completely fair scheduler~\cite{linux_cfs} 
for prompt inference. 
We design a new algorithm to partition the batch between 
prompts in different inference phases to achieve fairness while 
retaining throughput optimizations~\cite{sarathi_serve}.
Due to \sysname-enabled fast access to spare memory, the 
overhead of paging is minimal. Fair scheduling (1) achieves 
fair allocation of compute resources to all prompts, and 
(2) allows the model to stay responsive even under 
high load and limited GPU memory. We note that fair scheduling 
can absorb temporary bursts of queries or manage sustained
bursts until autoscaling can spawn new instances of the model.

\myparab{Results.}
We evaluate \sysname by hosting several state-of-the-art 
large generative ML models of different modalities (\eg text, 
audio, vision) on a server with 8 cutting-edge Nvidia H100 
80G GPUs. Using a representative distribution of these ML models
and inference workloads, we show that \sysname improves 
the responsiveness of LLM inference, measured using 
time-to-first-token, by $20\times$ compared to the state-of-the-art. 
\sysname improves the inference throughput over a 
single long prompt by $4\times$ on memory constrained inference~\cite{flexgen,hetegen2024}.


\section{Background and motivation}
\label{sec:background}
ML inference tasks are now critical for many organizations~\cite{openai_customer_stories}. End users often host ML models on cloud servers, issuing queries without deploying their own hardware~\cite{amazon-sagemaker,azure-ai-studio}. Generative models can produce various media, including text~\cite{llama}, images~\cite{stablediffusion}, and audio~\cite{audiogen}. LLMs, like ChatGPT, are particularly notable for their text generation capabilities.

\subsection{Infrastructure for ML}
At its core, ML inference entails 
performing several large matrix multiplications. Therefore, ML 
inference uses special-purpose accelerators like GPUs and TPUs, 
optimized for fast matrix multiplications~\cite{googletpuv4}. In addition 
to the number of floating point operations (FLOPs) supported by an 
ML accelerator, the memory on the accelerator has become 
crucial in enabling fast ML training and inference due to 
the rapid growth in the size of ML models. Generative ML models have
trillions of parameters, resulting in a large memory footprint.
To meet the memory footprint of models, GPUs are equipped with 
tens of gigabytes of \emph{high bandwidth 
memory} or HBM. Typically, GPUs are peripherals
connected to a host server using PCIe. The host machine has a 
CPU and DRAM for general-purpose computation~\cite{a100_dgx,b200_dgx,h100_dgx}.

\myparab{Server targets for ML.}
Multi-accelerator servers (\eg Nvidia DGX~\cite{dgx}, 
Cerebras WSE~\cite{cerebras}, Intel Gaudi~\cite{gaudi}) have become 
building blocks of larger cluster and datacenter deployments 
used for ML training and inference. A 
multi-accelerator server consists of a handful of ML accelerators,
connected with high-speed on-board electrical interconnects 
(\eg Nvidia NVlink~cite{nvlink}, Google ICI~\cite{googletpuv4}) 
that support up to hundreds of gigabytes per second of bandwidth 
between pairs of accelerators. Cloud operators connect racks 
of multi-accelerator servers into datacenter-scale deployments 
using a network fabric~\cite{googletpuv4,jupiter-evolving}.

\myparab{Bandwidth asymmetry in ML clusters.}
The interconnect between GPUs on the same server 
is specialized for high bandwidth at several hundred 
gigabytes/second~\cite{nvlink}, while the 
connection between GPUs and the host (\eg CPU, DRAM) relies 
on general-purpose PCIe (50GB/s), which is significantly slower
than \nvlinks in modern deployments. 

\subsection{Memory contention during LLM inference}
LLMs process inference queries, or \emph{prompts}, by converting them into tokens, which represent subword units. Each token is encoded as a multi-dimensional embedding vector that captures its meaning~\cite{word2vec,bert}. 
During inference, these token embeddings are fed into the LLM. 
Given a user prompt, the LLM predicts the most probable next token in the vocabulary~\cite{prob_gen} via the attention mechanism~\cite{attention}, which computes \emph{key} and \emph{value} vectors for each token by multiplying its embedding with learned key and value matrices~\cite{word2vec,bert}. Once the most likely token is generated, it is appended to the sequence, which includes both the prompt and previously generated tokens, and the process repeats. 

The key and value vectors of all tokens in this sequence, known as the inference context, are cached for efficient future predictions. The key and value vectors of simultaneously processed sequences are stored in a key-value (KV) cache within the GPU memory~\cite{vllm}. The KV cache size is determined by the sequence length and number of sequences, growing with the number of generated tokens. The KV cache size is limited by the available GPU memory after loading the model weights on the GPU. Memory contention arises when the KV cache exceeds the available GPU memory.

In fact, even a single prompt can experience starvation during 
inference when the model is too large for the GPUs, or the 
combined memory of the prompt's inference context (KV cache) and model 
weights exceeds GPU memory. As model sizes and supported 
prompt lengths grow rapidly (\eg Llama-405B and Gemini's 
1M prompt), the problem will worsen. This has prompted recent 
efforts to page context from the GPU to DRAM and load it only when 
necessary~\cite{flexgen,hetegen2024,llama31_blog}. 

\myparab{Memory contention hinders caching optimizations.} 
Recently, LLM serving engines are caching the inference 
context of long common prefixes of prompts~\cite{google_gemini,anthropic_caching}. 
As incoming prompts consume memory, the 
engine  must discard the cached context from the 
GPU, rendering the optimization useless. Instead, the 
only choice the engine has is paging out the inference context to the DRAM or remote memory~\cite{cache_gen}.

\subsection{Existing ways of mitigating memory contention}

One approach towards managing memory contention during LLM 
inference is to horizontally scale the number of GPUs hosting 
the model when a burst of inference requests arrives~\cite{llumnix}. 
However, spawning model instances on more GPUs 
incurs startup delays of several minutes~\cite{llumnix,aws-autoscaling-warm-pools}. 
These delays limit the utility of the scale out approach to requests that are
already admitted and queued, as they will experience unresponsiveness until more resources 
are provisioned.

\myparab{Paging suffers from bandwidth bottleneck.} 
Paging resolves the issues arising from memory contention 
but suffers from poor performance because GPUs are idle while 
they wait for the paged data. The fundamental bottleneck 
in paging is the limited bandwidth between the GPU and the 
DRAM which is constrained by the PCIe bandwidth. 
{While higher PCIe bandwidth may mitigate this problem, research has shown that the growth rate of PCIe bandwidth has been slow~\cite{bw_wall}.
Thus, current inference serving engines that offload dynamic context to host DRAM must trade-off performance for responsiveness.
In fact, to demonstrate this,
we implement a completely fair scheduler in vLLM and find that 
while fair scheduling improves the time-to-first-token in
vLLM (orange line Figure~\ref{fig:cfs_mb_ttft}), it causes 
request completion time or the time it takes to finish 
inferring a prompt, to increase by $\approx50\%$ (orange line in Figure~\ref{fig:cfs_mb_rct}).} 

\section{\sysname Design}
When inference load increases, modern serving systems 
take significantly longer to respond because of GPU memory capacity contention. Figure~\ref{fig:cfs_mb} shows that batch processing prompts with admission control~\cite{vllm,vtc} queues prompts until the current
batch of prompts finishes execution on the GPU, leading to head-of-line blocking delays. 
Serving can be unresponsive even when the inference load has a single prompt, when the GPU memory is insufficient for the model or the KV cache and model weights together exceed the available GPU memory~\cite{flexgen}. \sysname ensures responsive inference under GPU memory capacity contention by reducing preemption overhead through fast paging and implementing a preemptive scheduler that uses fast paging.


{
\myparab{Fast preemption with \sysname.} 
Multi-accelerator servers in ML clusters are allocated to inference jobs executing a variety of ML models. The inference jobs have different GPU compute core, memory bandwidth and memory capacity utilization based on the model (\eg image vs text generation). GPUs with certain inference jobs and workloads are bottlenecked by compute with significant spare memory capacity, while other jobs are bottlenecked by memory capacity~(\cite{usher},\S\ref{sec:profile}). We leverage this insight to decouple the memory and compute allocation on GPUs for inference jobs. This enables us to allocate memory on GPU with spare memory to another job bottlenecked by memory capacity on a different GPU. This allows the memory-constrained job to access additional memory via a high-speed GPU-GPU interconnect, such as \nvlinks, which offers an order of magnitude higher bandwidth compared to PCIe.}


\sysname addresses bursts of inference requests in 
cloud-hosted LLMs by ensuring responsive inference 
(low time-to-first-token) and maintaining high throughput 
(low request completion time). \sysname achieves the goal of 
responsive and high throughput inference even under bursts 
by implementing preemptive scheduling with low paging overhead 
(\S\ref{sec:cfs}). \sysname's fast preemption naturally benefits 
memory constrained inference systems reliant on paging~\cite{flexgen,hetegen2024,llama31_blog} and also advanced caching optimizations~\cite{anthropic_caching,google_gemini}. 
In \S\ref{sec:evaluation}, we show that \sysname also improves the performance of these systems in
addition to achieving responsiveness. \sysname has three main components (Figure~\ref{fig:overall-design}):

\begin{figure}[h]
  \centering 
  \begin{subfigure}[b]{0.4\textwidth}
    \includegraphics[width=\textwidth]{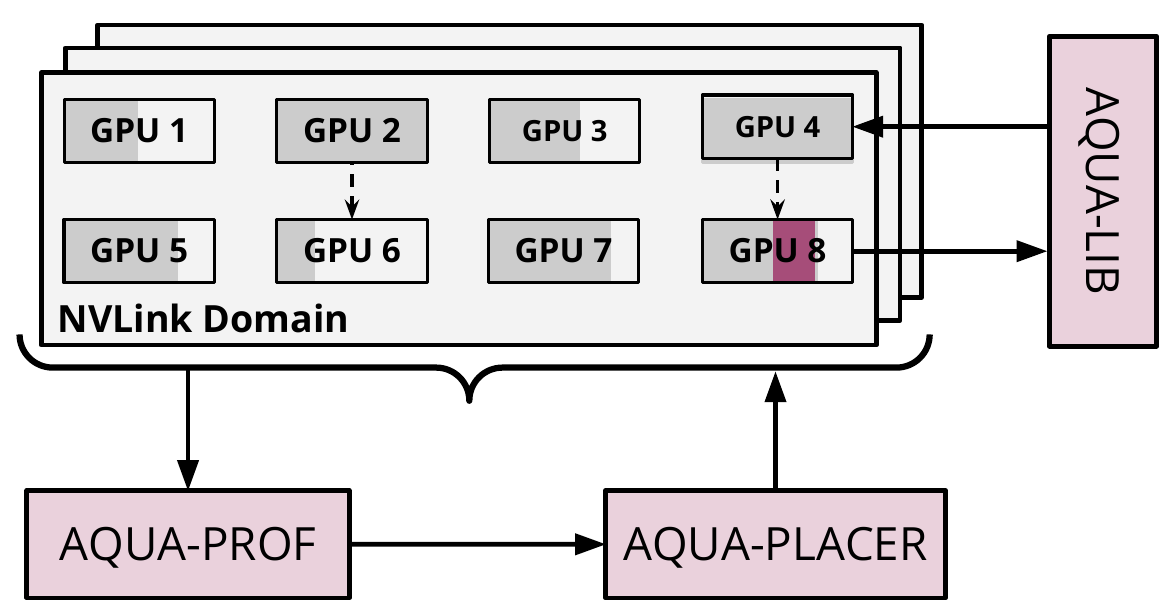}
  \end{subfigure}  
  \caption{\small{Design of \sysname. }}
  \vspace{-1em}
  \label{fig:overall-design}
\end{figure}

\myparab{1. \sysnameprofiler (\S\ref{sec:profile}).}
\sysnameprofiler profiles the characteristics of inference workloads
towards any hosted model and the corresponding GPU memory utilization.
Based on the profiled information, \sysnameprofiler decides if a GPU has excess 
memory or needs more GPU memory to sustain its inference load.
We term GPUs with excess memory as \emph{producers} 
and those that need memory as \emph{consumers}.

\myparab{2. \sysnamePl (\S\ref{sec:placement_algorithm}).}
Instead of relying on the chance existence of free GPU memory nearby, we 
develop an algorithm, \sysnamePl, to place models in clusters to 
maximize the opportunity to offload memory over the fast inter-GPU 
network. \sysnamePl places models on GPUs such that memory-bound ML 
models, \ie consumers, are hosted in proximity of memory-rich
ML models, \ie producers.

\myparab{3. \sysnamelib (\S\ref{sec:lib}).}
Inference workloads can change over time which means both the 
demand and supply of GPU memory during inference is elastic. 
This dynamic environment makes it hard to offload inference state 
to a different GPU that may require the memory back when it 
experiences high load of inference queries. To deal with this,
we develop \sysnamelib, a memory management framework that exposes
a new \sysnameT abstraction. \sysnameT are elastic --- 
they allow producer GPUs to transparently re-claim memory
they had exposed for offloads from memory consumer GPUs.


\section{\sysnameprofiler: profile memory consumption}
\label{sec:profile}
\sysnameprofiler profiles the current GPU utilization 
of inference jobs with the workload 
they serve in the steady state. Using this information, 
\sysnameprofiler estimates if the GPU is a producer or a consumer. {We call ML models on GPUs with spare memory capacity as producers and jobs that are bottlenecked by memory capacity as consumers.}


\begin{figure}[h]
    \centering
    \begin{subfigure}[b]{0.23\textwidth}
        \includegraphics[width=\columnwidth]{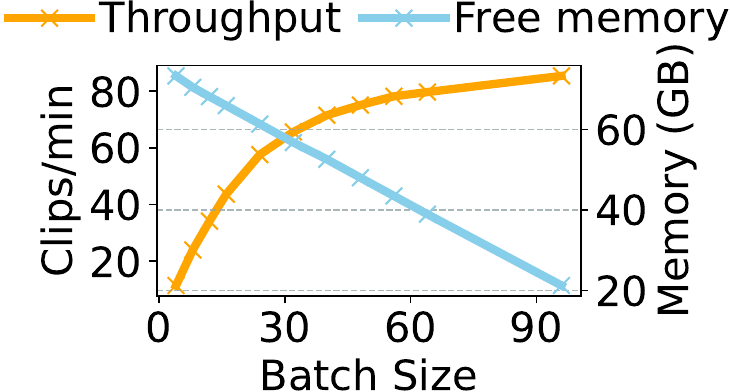}
        \caption{\small Audiogen}
    \label{fig:memory_avail_audio}
    \end{subfigure}
    \begin{subfigure}[b]{0.23\textwidth}
        \includegraphics[width=\columnwidth]{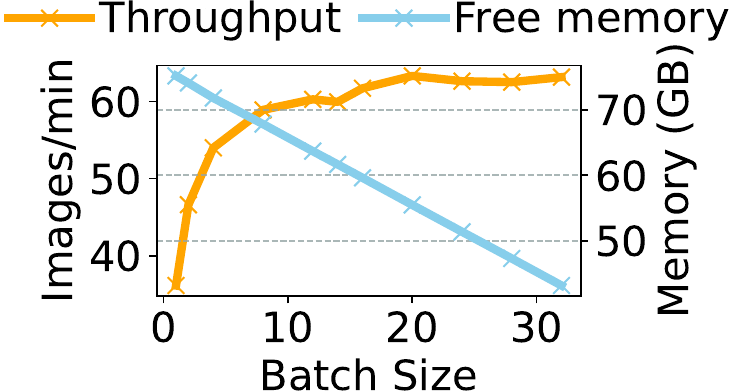}
        \caption{\small Stable diffusion}
    \label{fig:memory_avail_image}
    \end{subfigure}
    \caption{\small The plots show that when audio and image generation models reach the plateau of their throughput, there are 10s of GBs of free memory on an A100 80GB GPU. In fact, increasing the batch-size beyond a point results in diminishing increase in throughput.}
    \vspace{-1em}
    \label{fig:memory_throughput}
\end{figure}

Our experiments show that different generative ML models 
contend for different resources (\eg compute \vs memory). 
Specifically, vision and audio generative models have ample 
memory to spare even while operating at peak inference 
throughput (Figure~\ref{fig:memory_throughput}).{ Since we need to estimate the size of free memory at any time, \sysnameprofiler measures the free memory by increasing the batch size because 
the memory consumption of the model increases with batch size.}
We increase the batch size of inference until the model 
throughput saturates to measure the highest memory the model consumes.
Since partial responses in images and audio offer limited benefits, 
increasing the batch size further is unnecessary. 
We consider a model a producer if  
free memory at peak throughput on the GPU
is above a threshold (\eg 10 GB).

\subsection{Can LLMs be producers of memory?}
Our strategy of increasing inference batch size 
till throughput saturates is ineffective in 
determining the true memory requirement for LLMs. This is 
because LLM inference benefits from partial responses, 
especially in interactive use cases. Moreover, large 
batch sizes negatively impact the responsive time 
service-level agreements (SLA) of LLMs~\cite{sarathi_serve}. 
Therefore, we develop an alternate approach to classify LLMs.

\begin{figure}[h]
    \centering
    \begin{subfigure}[b]{0.22\textwidth}
        \includegraphics[width=\columnwidth]{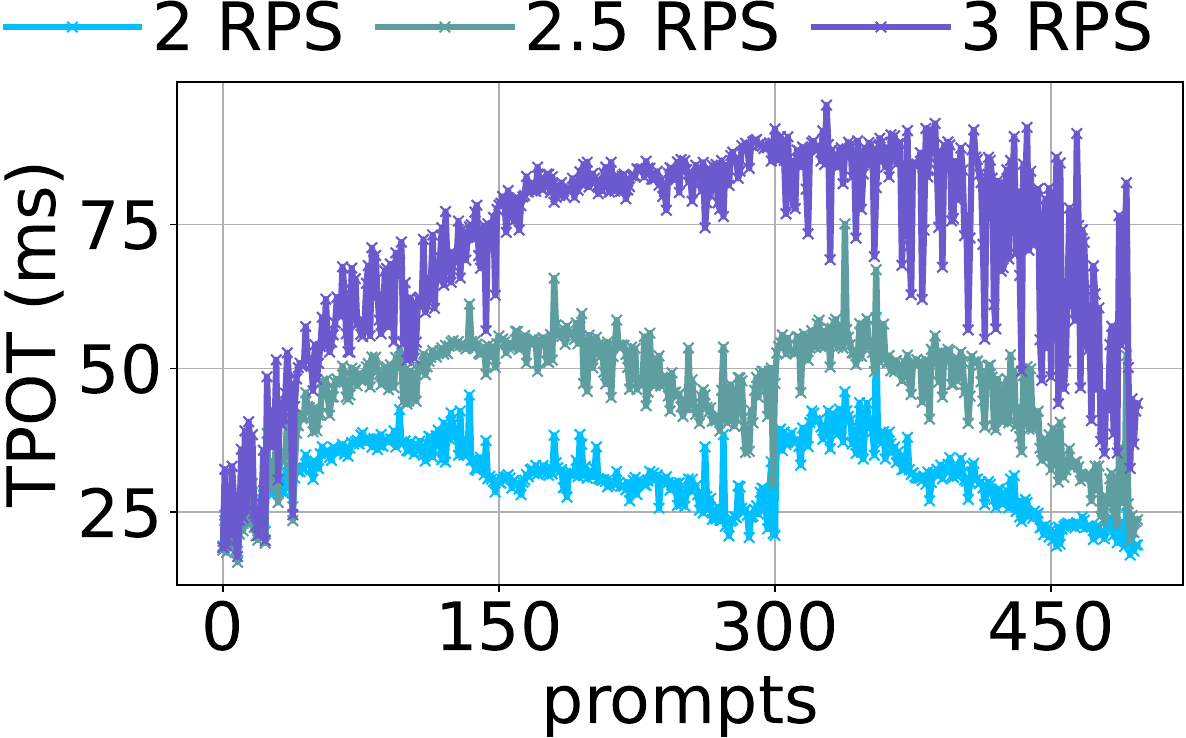}
        \caption{\small TPOT on ShareGPT}
    \label{fig:tpot_sharegpt_opp}
    \end{subfigure}
    \begin{subfigure}[b]{0.23\textwidth}
        \includegraphics[width=\columnwidth]{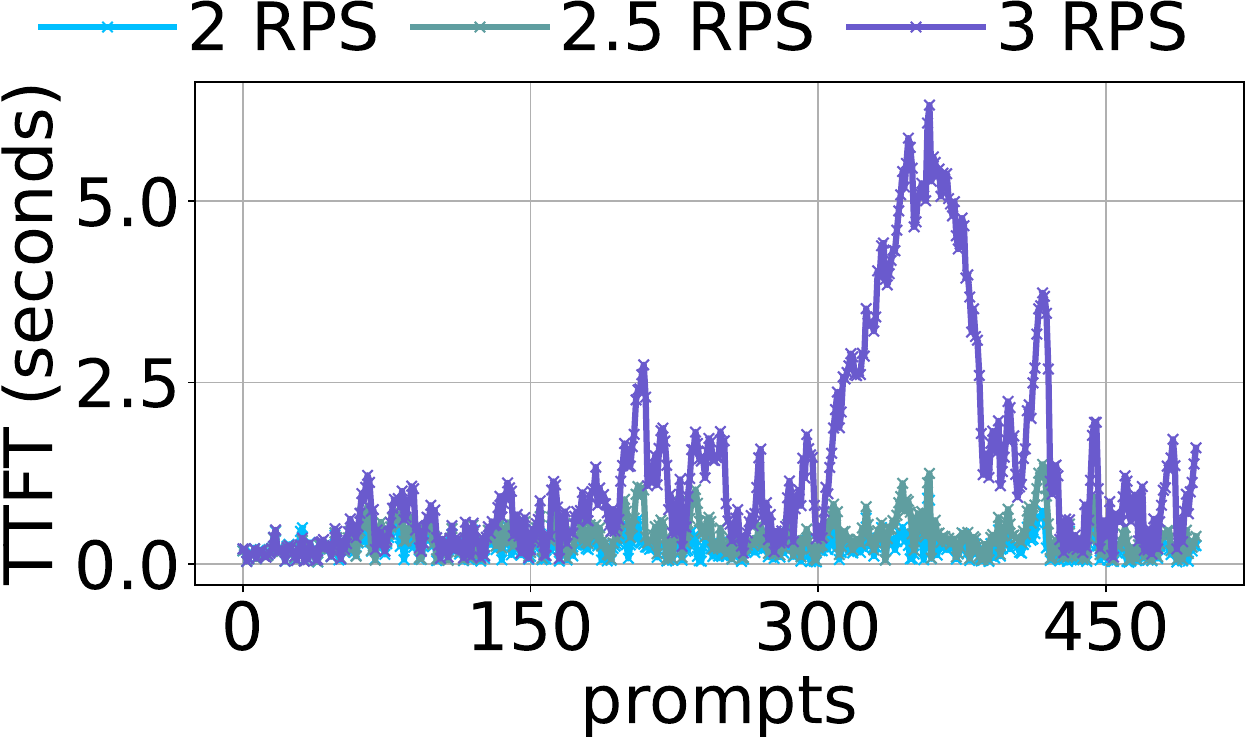}
        \caption{\small TTFT on ShareGPT}
    \label{fig:ttft_sharegpt_opp}
    \end{subfigure}
    \caption{\small \ref{fig:tpot_sharegpt_opp} shows the time per output token (TPOT) for the Llama 3.1 8B model on an 80 GB A100 GPU. TPOT increases with request rate. \ref{fig:ttft_sharegpt_opp} illustrates the time to first token (TTFT), rising with the queue length from incoming requests.}
    \label{fig:llm_opportunity}
\end{figure}

Inference on an LLM has two main stages: the prefill 
stage and the decoding stage. In the prefill stage, 
key and value vectors are calculated for each token in 
the prompt, and self-attention scores are computed for every 
token with all previous tokens~\cite{attention}. The output 
of the prefill stage is a vector per token, which is then 
multiplied by a matrix to generate the layer's output, 
preserving the input dimensions for the next layer. 
In the decoding stage, new tokens are generated with 
each forward pass. Unlike the prefill stage, where 
multiple vectors are processed, the decoding stage 
handles only one vector per prompt. 

Recent work has proposed inference schedulers that 
batch the prefill stage of some input prompts 
with the decode stage of other prompts for overall 
efficiency of inference~\cite{sarathi_serve}. 
The optimal batch size, or chunk size, depends on 
the required SLA of the hosted LLM.
Serving engines batch as many decode 
tokens as possible, supplementing the batch
with tokens from new prompts when necessary.


We observe that GPUs hosting LLMs often have free memory due to strict responsiveness SLAs, as batching is limited by chunk size. We empirically demonstrate this using SLA requirements for interactive use cases from recent work~\cite{sarathi_serve}, specifying a 100ms P99 Time Between Tokens (TBT) and under 1 second Time to First Token (TTFT). We also adopt their recommended chunk size of 512 for the ShareGPT~\cite{vllm} dataset to meet these SLAs.

\myparab{Challenge.} 
Given the SLA, chunk size, and dataset, \sysnameprofiler 
searches for the highest request rate that meets the SLA
since it leads to maximum memory consumption. 
Sweeping the entire request rate space is infeasible and challenging. 
To overcome this, we first perform a binary search within a range (\eg 1-10 RPS) followed by a linear search with 0.5 RPS increments and decrements to find local maxima.
Using this, \sysnameprofiler 
outputs the available memory at the highest traffic rate the LLM can sustain within
the SLA. 

In Figure~\ref{fig:llm_opportunity}, the Llama 3.1 8B model handles 2.5 RPS without violating the TTFT SLA. At 3 RPS, the queue size increases, resulting in a TTFT of 5-6 seconds (Figure~\ref{fig:ttft_sharegpt_opp}), which is unacceptable. Thus, \sysnameprofiler reports the available memory (40 GB) at 2.5 RPS as free memory for this model and if this value exceeds a threshold (e.g., 10 GB), we classify the model as a producer. 

\myparab{Unpredictable loads.}
{Unlike image generation models where the memory consumption of every data point in a batch is the same, the memory consumption of each prompt in a batch for LLM inference can vary significantly based on the length of the input and output tokens. \sysnameprofiler samples requests for profiling using input and output token distributions from individual LLM deployments~\cite{AzureLLMInferenceDataset2024}. Its free memory estimation remains accurate as long as request patterns follow the historical distribution.} Even with extensive profiling, a GPU's memory utilization fluctuates 
with unpredictable workload changes. To handle this, \sysname swiftly reclaims memory and returns it to the producer (\S~\ref{sec:lib}).

\myparab{Profling consumers.}
If the model is not a producer, \sysnameprofiler assesses its need for paging and preemption. {If the model has allocated swap space using \sysnamelib}, the model is classified as a consumer, and \sysnameprofiler determines the required swap space. For memory-constrained inference and prefill caching, this space is computed based on the number of prompts, the longest sequence required, and the memory needed for inference context. \sysnameprofiler empirically determines the swap space required for CFS through artificial bursts of varying sizes (e.g., 2X, 3X the steady-state requests per second) (\S~\ref{subsec:adaptability}). If there is no need for swap space, we treat the model as a producer with 0 memory to offer.

{
\myparab{Corner case performance.} In corner cases, when all models are consumers of memory, \sysname defaults to using DRAM for paging, matching the performance of baselines like vLLM and FlexGen since they page to DRAM over PCIe.
}

{
\subsection{\sysname in future datacenters}
Recent trends in ML infrastructure suggest that \sysname's techniques are well-suited for modern datacenters. \sysname targets environments where multiple ML jobs co-exist within high-bandwidth interconnect domains like the \nvlinks/ NVSwitch fabrics. These domains are growing in size, with examples like Nvidia's NVL72 racks (72 GPUs interconnected with NVSwitch)~\cite{nvidia_blackwell_2024} and Google's TPU racks (64 TPUs per block)~\cite{googletpuv4,hotnets_photonics}. The large size of these scale-up architectures increases the likelihood of diverse ML workloads co-existing within the same high-bandwidth domain. Diverse workloads enable \sysname to match a higher number of producers with consumers compared to smaller multi-GPU servers.
}

\section{\sysnamePl: optimal model placement}
\label{sec:placement_algorithm}

\sysnamelib's ability to offload tensors on 
interconnected GPUs relies on the availability of free 
memory on these GPUs. \sysnamePl maps ML models for inference to GPUs in a cluster
of servers, with the goal of maximizing opportunities to 
offload \sysnameT on interconnected GPUs. 
Mapping a single producer to multiple 
consumers is feasible but \sysnamePl does not allow that 
by design and maps one producer to one consumer. Sharing a producer with 
multiple consumers may oversubscribe NVlink bandwidth of 
the producer GPU, reducing the benefits of offloading memory over 
the inter-GPU network. 


\myparab{Differences from stable matching.}
Modeling the placement as a simple stable 
matching formulation will assume that any producer can be 
matched with any consumer. However, in the context of \sysname, 
we can only pair producers to consumers that are 
connected by the same fast server-scale 
inter-GPU network (\eg NVlinks or ICI).
\sysnamePl solves the model assignment problem in two steps. 
First, it assigns models to multi-GPU servers and then within 
each server, it matches producers to consumers using 
simple stable matching. We encode the first step as an 
optimization to produce optimal model to server mappings.

\myparab{Inputs and outputs.} The optimization  
takes as input the number of servers in the cluster $S$, 
number of GPUs per server $G$, the number of models $N$ and 
the total high-bandwidth memory capacity of each GPU $G_{mem}$. 
We make a simplifying assumption that all GPUs in a server 
have the same hardware specifications (\eg compute cores 
and memory) inspired from  
the current generation multi-GPU servers~\cite{dgx,b200_dgx}.
A key input is the memory 
requirement of each model $R_m$, which we determine using \sysnameprofiler. 
The model's memory requirement 
is positive if it is a producer and negative if it is a 
consumer. For example, Audiogen in Figure~\ref{fig:memory_avail_audio} 
will have $R_m = 20 GB$. The output of the encoding is 
an indicator variable $x_{m,s}$ that is $1$ if model $m$ 
is mapped to server $s$. To handle distributed serving, we 
assign indicator variables $x_{i,s}$ to each model
shard $i$. $P$ is a list of mappings, each storing a set of all the
tensor parallel shards of a model. 

Allocation of models to GPUs is subject to the constraints: 

\setlength{\abovedisplayskip}{1pt}
\setlength{\belowdisplayskip}{1pt}

\myparab{One model per GPU.} Each model must map to one server: 
\begin{equation}
    \sum_{s=1}^{S} x_{m,s} = 1, \quad \forall m \in M
    \label{eq:one_gpu_per_model}
\end{equation}

\myparab{Tensor parallel shards are on the same server.}
Tensor parallel (TP) shards are typically placed on the same server and pipeline parallel shards on different servers~\cite{vllm_dist,sarathi_serve}. Accordingly, we map all TP shards of a model to a single server.
\begin{equation}
    x_{p[j],s} == x_{p[j+1],s} \quad \forall j \in \{0 \ldots len(p) - 1\}, \forall p \in P
\end{equation}

\myparab{Models limited by number of GPUs per server.} 
In \sysname, an ML model shard is placed on exactly one GPU.
If one shard maps to a single GPU, the number of shards 
mapped to a server should not exceed the number of GPUs 
on the server:
\setlength{\abovedisplayskip}{1pt}
\setlength{\belowdisplayskip}{1pt}
\begin{equation}
    \sum_{m=1}^{M} x_{m,s} \leq G \quad \forall s \in \{1 \ldots S\}
    \label{eq:one_model_per_gpu}
\end{equation}

\myparab{Balanced demand and supply of memory.}
We define $mem_{s}$ as the sum of the memory 
requirements ($R_m$) of models mapped to a server. 
Recall that $R_m$ can be positive (for a consumer) or 
negative (for a producer). Thus, $mem_{s}$
captures the net memory available on the server after mapping 
models to it. 
$mem_{s}$ will be positive if there is 
excess memory or negative if there is a deficit.

\setlength{\abovedisplayskip}{1pt}
\setlength{\belowdisplayskip}{1pt}
\begin{equation}
    mem_{s} = \sum_{m=1}^{M} x_{m,s} \cdot R_{m} \quad \forall s \in \{1 \ldots S\}
    \label{eq:server_util}
\end{equation}

\myparab{Balanced number of consumer and producers.}
Despite keeping the memory usage on 
the server balanced using equation~\ref{eq:server_util},
the encoding can find solutions 
that map one producer and multiple consumers to the same 
server. We count the 
net number of models mapped to a server, where a consumer 
counts for $-1$ and a producer counts for $+1$ in Eq~\ref{eq:server_eq}.
We find solutions where $eq_{s}$ is close to 0 to 
prevent producer bandwidth oversubscription.

\setlength{\abovedisplayskip}{1pt}
\setlength{\belowdisplayskip}{1pt}
\begin{equation}
    eq_{s} =  \sum_{m=1}^{M} x_{m,s} \cdot t \quad \forall s \in \{1 \ldots S\}, \quad t = \frac{|R_{m}|}{R_{m}}
    \label{eq:server_eq}
\end{equation}

\myparab{Objective.}
\sysnamePl's goal is to ensure that models 
are mapped to servers such that there is no waste of 
GPU memory across servers (Eq~\ref{eq:server_util}) and 
each server has a balanced number of consumers and producers
(Eq~\ref{eq:server_eq}). We achieve this by picking the 
max of both memory utilization ($mem_s$) and number of models
($eq_s$) on a server and minimize their largest sum.
To make the two parts of the objective have the same 
units, we multiply $eq_s$ with the server's GPU memory:
\setlength{\abovedisplayskip}{1pt}
\setlength{\belowdisplayskip}{1pt}
\begin{equation}
    \max_s(mem_s) + G_s \cdot \max_s(eq_s)
    \label{eq:objective}
\end{equation}

We encode the algorithm
using a commercial optimization solver, Gurobi~\cite{gurobi}
and find that it reaches solutions within 2\% of optimal
in 5 seconds for clusters with up to 128 GPUs.

\myparab{How frequently do we run \sysnamePl?}
When producer workloads change, \sysnamelib rapidly reclaims memory for the producer, limiting consumers' access to fast paging. While this prevents performance degradation for producers,  the state of the cluster itself might not be optimal to maximize offloading opportunities. Infrastructure providers must thus trade off between maintaining suboptimal offloading or rerunning \sysnamePl to remap jobs with migration for better sharing opportunities. This trade-off is not explored in this paper and is left for future work.
\section{\sysnamelib: elastic memory management}
\label{sec:lib}
\begin{figure}[h]
    \centering 
    \begin{subfigure}[b]{0.45\textwidth}
      \includegraphics[width=\textwidth]{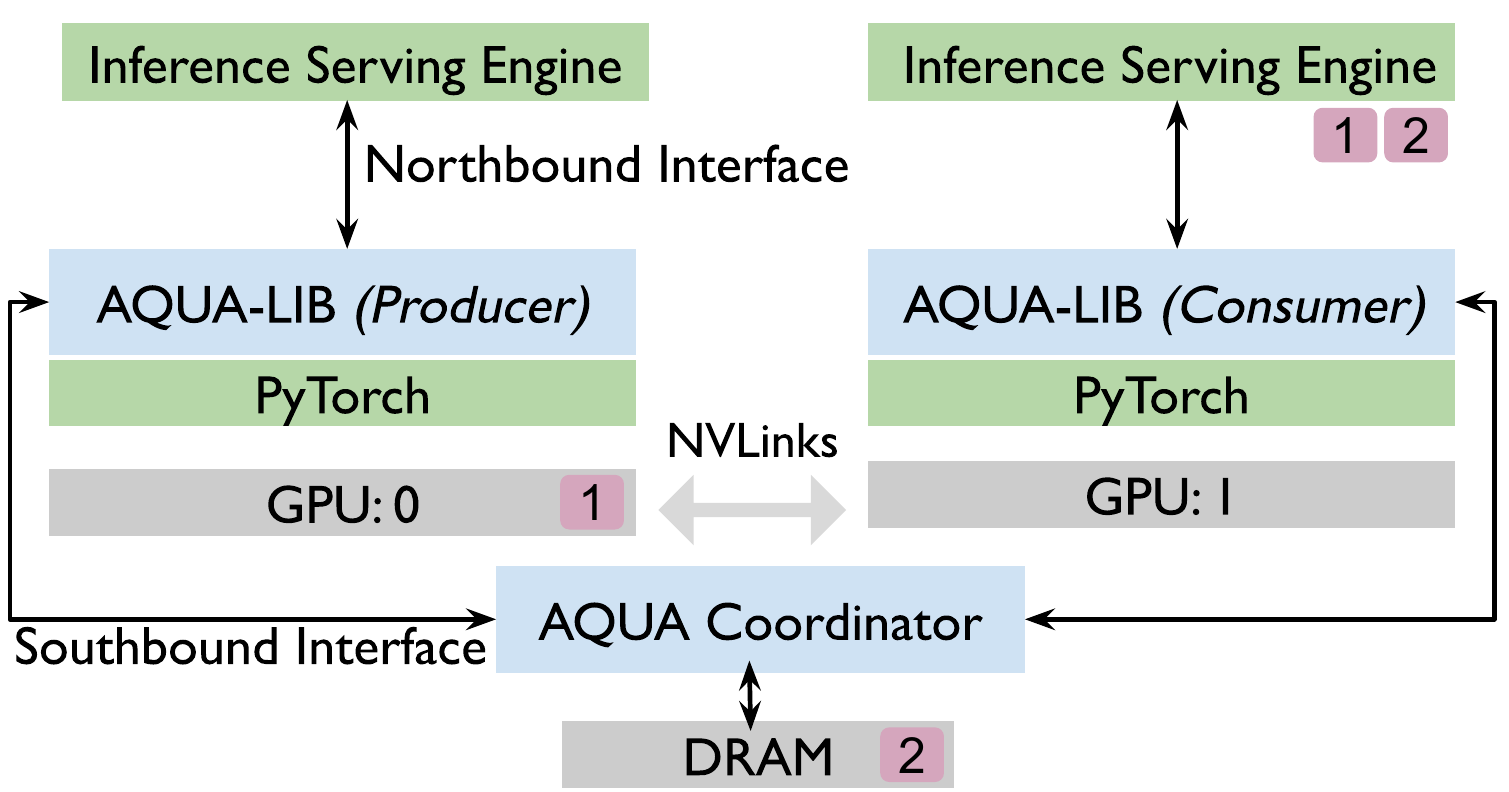}
    \end{subfigure}     
    \caption{\small{Design of \sysname. GPU 1 is hosting 
    a consumer. \sysnamelib
    is aware that GPU 0 is a producer. 
    \sysnamelib allows the model on GPU 1 to allocate 
    \sysnameT that are offloaded to GPU 0's HBM (shown 
    in a pink box with number 1). If GPU 0 only 
    has enough memory to offload one tensor, \sysnamelib
    falls back to the host DRAM (
    shown using a pink box with number 2).}}
    \label{fig:aqua_design}
\end{figure}

\sysnamelib implements a novel
abstraction for \emph{migratable} tensors in GPU memory, 
called \sysnameT. 
\sysnameT can be offloaded on to the memory of other 
GPUs using the fast inter-GPU network. Internally, 
\sysnamelib decides when \sysnameT should be migrated 
and to which GPU's memory, based on the memory 
utilization and inference workload on all other GPUs 
in the server. The ML model developer is oblivious to 
changes in the physical location of \sysnameT.
We integrate \sysname with two popular ML serving engines, 
vLLM~\cite{vllm} and FlexGen~\cite{flexgen}, and the integration does not need big 
changes to their codebase (\S\ref{sec:implementation}).

\myparab{\sysnamelib's interfaces.}
An instance of \sysnamelib runs on each GPU of a multi-GPU 
server (Figure~\ref{fig:aqua_design}). \sysnamelib is 
initialized with the profile generated by \sysnameprofiler.
\sysnamelib uses this information to decide if the 
GPU has spare memory to be a producer. \sysnamelib implements a 
\emph{northbound} and a \emph{southbound} interface.
Serving engines use northbound interface to interact with \sysnamelib. 
Using the northbound interface, serving systems also
share their metrics like inference load (e.g., requests per second) and size of inference context with \sysnamelib at a tunable frequency.  
If the GPU is a producer and the serving engine reports an 
increase in inference load, \sysnamelib will
re-claim the offloaded memory. \sysnamelib will 
offer the memory back to consumer when the load returns to the steady
state value determined during profiling by \sysnameprofiler. When \sysname 
offers memory from a producer or reclaims it back, 
it uses the northbound interface to inform the inference 
serving engine of how much memory it has after the event.
The southbound interface enables \sysnamelib to 
communicate with a centralized thread-safe data store maintained 
by the \sysname \emph{coordinator}. The coordinator tracks 
\emph{requests} for memory offloads from consumer GPUs 
and memory \emph{offers} from producer GPUs. 


\myparab{Central coordinator.}
The central coordinator keeps track of \emph{consumers} and
\emph{producers} of HBM. 
Consumer GPUs need to offload tensors 
to the HBM memory of any potential producer GPUs. 
The coordinator program exposes a set of REST endpoints.
\sysnamelib uses these REST endpoints to register requests
for memory from consumer GPUs and offers of free memory 
from producer GPUs. The coordinator also exposes REST 
endpoints that allow producer GPUs to re-claim memory 
when they need it back. A GPU may 
start as a memory producer, but it may reclaim all the offered memory
if the characteristics of its inference workload change.


\myparab{Allocating \sysnameT.}
An ML model imports \sysnamelib and initializes 
it with the profile generated by \sysnameprofiler.
ML models use \sysnamelib to instantiate \sysnameT. 
\sysnamelib uses the southbound interface to register 
this allocation request with the coordinator via the appropriate REST 
endpoint (\S\ref{sec:implementation}). The coordinator 
keeps track of GPUs that may have registered to be 
producers. Selecting which GPU will be the producer for 
a consumer GPU is explicitly done by the \sysnamePl 
(\S\ref{sec:placement_algorithm}) before the model starts 
to execute on the consumer GPU.
The coordinator returns a reference of the offloaded tensor 
location to the consumer's \sysnamelib. We note that 
if no producer GPUs exist in the system, \sysnamelib falls 
back to using the DRAM for offloading tensors, just like previous 
work~\cite{flexgen, vllm}. Freeing allocated tensors 
follows a similar logic where the consumer's \sysnamelib 
instance informs the coordinator which centrally manages 
bookkeeping of \sysnameT.

\myparab{Reclaiming \sysnameT.}
On producer GPUs, \sysnamelib polls the northbound 
interface to find the current memory utilization 
and workload statistics of the GPU. 
Using this, \sysnamelib allows producers 
to reclaim their memory back from offloads when the 
inference load on the producer GPUs increases. This 
occurs when \sysnamelib finds that the workload 
reported by the inference serving engine has increased.
\sysnamelib uses the REST API to let the coordinator 
know that the producer GPU would like the memory back. 

\myparab{\sysnamelib control loop.}
On a consumer GPU, \sysnamelib's control logic tests 
once every few inference iterations if any offloaded \sysnameT
need to be freed up. Doing this check requires \sysnamelib 
calling the coordinator's REST API. This check ensures that 
if a producer has registered a request to reclaim 
its memory, consumers can respond to it in a timely manner.
We note that the centralized coordinator's data store 
is thread-safe to prevent unexpected behavior when 
multiple GPUs make changes. In our design 
the overheads of \sysnamelib are very low 
since communication with central coordinator is 
infrequent -- only once per a configurable number 
of inference iterations. We describe further
implementation details of \sysnamelib in 
\S\ref{sec:implementation}.
\section{Fair scheduling prompts with \sysname}
\label{sec:cfs}

Default behavior of the scheduler in today's inference serving
systems is to admit new requests only if there is 
enough GPU memory available for its inference context~\cite{vllm,sarathi_serve}. 
Schedulers like VTC~\cite{vtc} provide fairness at a coarse granularity
with admission control to avoid users with a high query rate consuming 
all the compute cycles. 
However, such coarse grained admission control can 
still lead to starvation when there is a surge in the number of legitimate users,
beyond the serving capacity of the allocated hardware.
If the GPU memory is fully occupied by prompts already being 
inferred, any additional requests need to \emph{wait} or
\emph{starve} until enough space becomes available in the 
GPU memory. Even with auto-scaling, prompts can starve because 
provisioning new VMs incurs delays in the order of minutes~\cite{aws-autoscaling-warm-pools}.


\begin{figure}[h]
    \centering 
    \begin{subfigure}[b]{0.48\textwidth}
      \includegraphics[width=\textwidth]{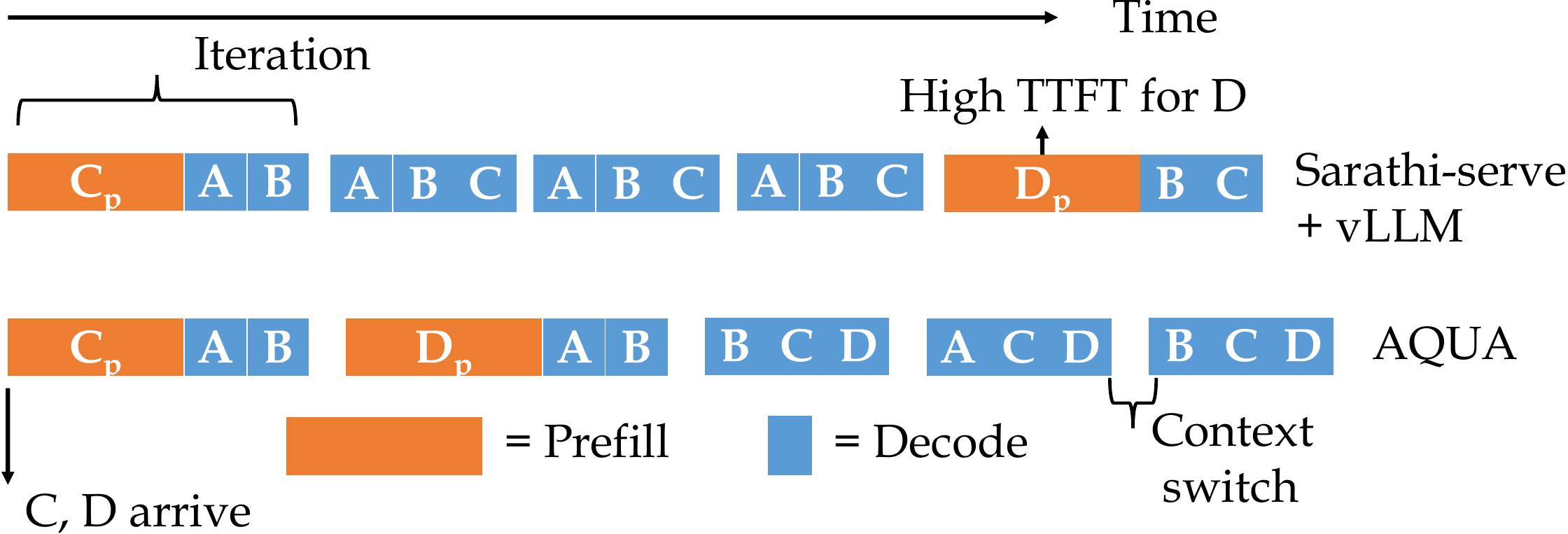}
    \end{subfigure}    
    \caption{\small{CFS enabled by \sysname during memory contention where only 3 out of the 5 prompts can fit on a GPU at a time.}}
    \label{fig:cfs_design}
    \vspace{-1em}
\end{figure}

\myparab{Batching \vs timesharing for scheduling inference.} 
Recall that state-of-the-art scheduling algorithms batch chunks of prefill computation
with decoding computation of other prompts to increase compute efficiency (\S~\ref{sec:profile}).
The algorithm prioritizes decodes while batching prompts, which then leads to  
starvation of a subset of prompts when multiple prompts arrive simultaneously. 
In, Figure~\ref{fig:cfs_design}, C,D, 
arrive at the same time but D starves till A completes execution because the GPU's memory can accommodate only 3 prompts at a time. 
This contrasts with the behavior of operating systems
that use context-switching to manage constrained resources under load with a completely fair scheduler (CFS)~\cite{linux_cfs}. 
Context-switching has two steps: \emph{preemption}, 
which saves the current process's state 
(registers, page table, \etc) to DRAM, and \emph{loading},
which loads the next process's state into the CPU.

\myparab{Challenge.} 
Naively applying time-slicing from OS scheduling, which prioritizes processes that have received the least compute time, undermines recent chunked-prefill optimizations~\cite{sarathi_serve}. These optimizations leverage the fact that prefill is compute-bound while decoding is memory-bandwidth-bound, allowing prefill and decoding tokens to be batched together, effectively providing "free" decoding. Time-sharing GPUs based on the least number of generated tokens would only batch prefill prompts, as they have zero generated tokens, negating this benefit.

\myparab{\sysname's batch partitioning algorithm.}
In \sysname, we achieve fair scheduling while preserving the benefits of chunked-prefill. Given a batch size $b$, we partition it into $p$ prefill tokens and $d$ decode tokens. Our key insight to partition is to set $d$ to its upper bound, corresponding to the maximum number of prompts that fit within the GPU memory. \sysname first fills $p$ with prefill prompts having the least number of prefill tokens computed, $d$ with decode prompts with the least number of tokens generated. The remaining slots in $d$ are allocated to prompts in $p$. This ensures full utilization of GPU capacity when no prefills are present, while leveraging chunked-prefill optimization when they are. \sysname stops filling the batch if the GPU's memory is exhausted.

For simplicity, we measure time slices as the number of inference iterations, as each iteration of a given batch size takes the same time~\cite{gandiva}. \sysname reschedules the batch every $k$ iterations or when a request completes, paging out prompts that
are not a part of the next batch and paging in prompts that were not on the GPU.
Figure~\ref{fig:cfs_design} shows how \sysname fairly allocates compute between prefill and decoding prompts, reducing TTFT for prompt D.

\myparab{Efficient context switching.} 
Simply using \sysnameT as swap space is suboptimal for two reasons.
First, vLLM stores the key-value tensors of all the prompts associated with a layer as one 
tensor. So, a given token's key and value tensors are 
scattered across tensors of different layers that are discontiguous 
and this leads to multiple 
small copies to the offloaded \sysnameT. Second, \nvlinks bandwidth is poor for 
small data transfers, \eg between two A100 GPUs, the copy bandwidth 
of a 4MB buffer is 50 GB/s and 64 MB buffer is 200 GB/s. We solved 
this challenge by gathering smaller tensors into a 
temporary tensor on the GPU and copying that to the 
offloaded tensor. Similarly, \sysnamelib
copies offloaded data to a temporary tensor, and
scatters it to respective smaller tensors. 

{
  Finally, the serving engine can query \sysnamelib for the tensor location and fall back to FCFS from CFS when tensors are on DRAM to avoid high context switching overheads.
}

{
  \myparab{Other scheduling policies.} Note that any policy that requires swapping can be implemented using \sysname for faster performance compared to naively swapping to DRAM. For example, priority fairness can be implemented with \sysname by preempting lower priority to swap space hosted with \sysnameT. \sysnamelib places these tensors on an interconnected GPU's memory whenever possible.
}
\section{Implementation}
\label{sec:implementation}
We implement \sysnamelib as a pip-installable Python library utilizing PyTorch~\cite{pytorch} tensors with 500 lines of code (LOC). The library provides functions to convert standard CPU-based tensors to \sysnameT and retrieve the corresponding GPU or DRAM pointer for compute access. Internally, \sysnamelib communicates with the coordinator. We integrate the library's northbound interface with serving engines, allowing users to specify GPU memory offering and retention, managed by \sysnameprofiler.

In vLLM v0.5.3, we implement CFS (\S\ref{sec:cfs}) with 150 LOC, alongside a CUDA kernel to efficiently gather small buffers into larger ones for \nvlinks transfer with 200 LOC. We introduce a new prefill caching API with unique IDs, and both CFS and prefill caching share the same swap space for context preemption. \sysnamelib is integrated into vLLM and FlexGen~\cite{flexgen} to manage swap space using \sysnameT instead of PyTorch CPU tensors in less than 50 LOC. A key challenge for \sysnameT is moving tensors from GPU to DRAM without causing data corruption. We solve this via the insight that inference engines have execution loop boundaries with no GPU compute, allowing safe transfers.

We implement \sysnamePl using Gurobi~\cite{gurobi} in 700 LOC. \sysname coordinator is a socket server with REST APIs to offer, reclaim and allocate, deallocate memory as producers and consumers respectively, implemented in 250 LOC.
\section{Evaluation}
\label{sec:evaluation}
In this section we demonstrate that \sysname's fast preemption significantly improves responsiveness and throughput.

\myparab{Experiment testbed.}
Our experiments utilize two separate testbeds. The primary testbed features a server with eight NVIDIA H100 GPUs, each with 80 GB of memory. These GPUs are interconnected via \nvlinks and NVSwitches, providing 450 GB/s bandwidth, and connected to DRAM via PCIe Gen5 with 50 GB/s bandwidth and the server has 1.5 TB of RAM. The secondary testbed consists of a server with two NVIDIA A100 GPUs, each with 80 GB of memory, directly connected via \nvlinks with 300 GB/s bandwidth. These GPUs connect to DRAM via PCIe Gen4 with 25 GB/s bandwidth. We use the H100 testbed for end-to-end tests and the A100 server for micro-benchmarks to minimize costs.

\subsection{Models and workloads.}


\myparab{Datasets.}
We use the parti-prompts~\cite{parti_prompts} dataset to evaluate vision generative models and the default prompts from Audiogen to evaluate audio generative models. Following recent research~\cite{sarathi_serve}, we use the sharegpt~\cite{vllm} and arxiv summarization~\cite{arxiv-summary} datasets to evaluate LLMs. 

\myparab{Distributed serving.}
We adopt the default policy in vLLM of using tensor parallelism to distribute models across GPUs during inference~\cite{vllm,megatron} as long as the model fits within a server and vary the number of GPUs a model is split across to represent diverse workloads. Specifically, Llama 70B can be deployed on 2 GPUs using tensor parallelism, but it can also be scaled to a larger number of GPUs at higher cost when traffic demands are consistently high. For example, when running on the ShareGPT dataset with H100 GPUs, increasing from 2 to 4 GPUs enables Llama 70B to handle ten times more requests, as 4 GPUs provide ample memory after loading the model weights. However, if an organization expects lower sustained traffic with occasional bursts, it is more cost-effective to use fewer GPUs and handle bursts as they arise. In our evaluation, we study both configurations and allow \sysnameprofiler to dynamically determine when it should function as either a producer or consumer.

\myparab{Producer workloads.}
For audio and vision generative models, we execute the model with a batch size that reaches the peak throughput first. For Llama 8B and 70B models, we use the ShareGPT dataset and run it with interactive workload SLA~\cite{sarathi_serve} and pick a request rate that maintains a steady queue length, \ie the request arrival rate is equal to the request completion rate. We use the same chunk size as recent efforts that developed chunked prefill~\cite{sarathi_serve} in our experiments.

\myparab{Consumer workloads.}
We evaluate three different consumer workloads, a bursty workload that requires fair scheduling (CFS) introduced in \S~\ref{sec:cfs}, a prefill caching workload and a memory-constrained long prompt workload. 

\myparab{Fair scheduling during bursts.} 
We use two models to evaluate the bursty workload, Yi-34B on 1 GPU and Llama-70B on 2 GPUs distributed with tensor parallelism. We determine the stable request rate for both the models on the ShareGPT dataset and double the request rate for a duration of 1 minute to create a burst. Assuming an upper bound of 1 minute to provision new instances to handle the burst, we implement the state-of-the-art approach proposed in recent work~\cite{llumnix} to handle load-balancing while auto-scaling. We track the request arrival rate every 5 seconds and whenever the request rate increases, we trigger an auto-scale notification. Our workload generator listens to the auto-scale notification and  simulates the behavior of the load-balancer by first waiting for a minute for the VM to spin up and then redirecting the newly arriving requests to a dummy http server until the number of requests in the burst has been redirected. At this point, all the instances are back to serving requests at the stable rate. We use a dummy server instead of a VM to avoid allocating additional GPUs and save cost. 

\myparab{Prefill caching.} 
We use Yi-34B-200K model served using vLLM on 2 GPUs to have enough GPU memory to accommodate a large prompt and evaluate prefill caching. We pick a large prompt with 100 thousand tokens to cache and interleave this prompt along with the ShareGPT dataset. Each time the prompt is issued, we only modify the last 256 tokens to reuse the cached prompt to model use cases such as question answering on a long context.

\myparab{Long prompt.}
We use the OPT-30B model served using FlexGen~\cite{flexgen} to evaluate long prompts that exceed the available memory on the GPU after loading the model weights. FlexGen offloads the inference context of a prompt to the DRAM by preempting the previous layer's context and loading the next layer's context. We determine the median prompt length from the arxiv summarization dataset to be over 7000 tokens, so we sample prompts of length 8192 to evaluate FlexGen since its evaluation script requires a single prompt length parameter for the duration of the experiment.




\begin{table}[h]
    \centering
    \small
    \begin{tabularx}{\columnwidth}{X X X c c}
        \hline
        \textbf{Model} & \textbf{Workload} & \textbf{Engine} & \textbf{GPUs} & \textbf{Type} \\
        \hline
        Llama3.1 70B & ShareGPT & vLLM + CFS & 2 & C \\
        \hline
        Yi-34B-200K & ShareGPT + Prefill Cache & vLLM & 2 & C \\
        \hline
        OPT-30B & Long-prompt & FlexGen & 1 & C \\
        \hline
        Llama3.1 8B & ShareGPT & vLLM & 1 & P \\
        \hline
        Llama3.1 70B & ShareGPT & vLLM & 4 & P \\
        \hline
        SD-3, SD-XL, Kandinsky & Parti prompts & Diffusers & 1 & P \\
        \hline
        MusicGen, AudioGen & Audio descriptions & PyTorch & 1 & P \\
        \hline
    \end{tabularx}
    \caption{Models, workloads, serving engines, and their type (P = Producer, C = Consumer). SD = Stable-Diffusion.}
    \label{tab:models}
\end{table}

\subsection{Baselines}

We evaluate the CFS workload using two baselines: the default vLLM scheduler (FCFS) and vLLM with CFS. While non-preemptive admission control schedulers like VTC~\cite{vtc} prevent malicious users from monopolizing compute resources, they still suffer from queuing delays during a surge of legitimate users. To simulate this behavior, our vLLM baseline treats each request in a burst as a new unique user, capturing the unresponsiveness. Thus, the vLLM lines in our graphs also represent VTC performance.

We use our implementation of vLLM with prefill caching that preempts to DRAM as the baseline for prefill caching. We use the vanilla FlexGen preempting to DRAM as the baseline to evaluate long prompt workload. All experiments use the default scheduler unless specifically evaluating CFS.

\myparab{Evaluation metrics.}
We use Time To First Token (TTFT), Time Per Output Token (TPOT) and decoding throughput or generation throughput in tokens per second. 

\subsection{End to end evaluation.}

We run an end-to-end evaluation on 8 GPU H100 servers and we set the cluster size to 16 to avoid high costs. 

\begin{figure}[h]
    \centering
    \begin{subfigure}[b]{0.22\textwidth}
        \includegraphics[width=\columnwidth]{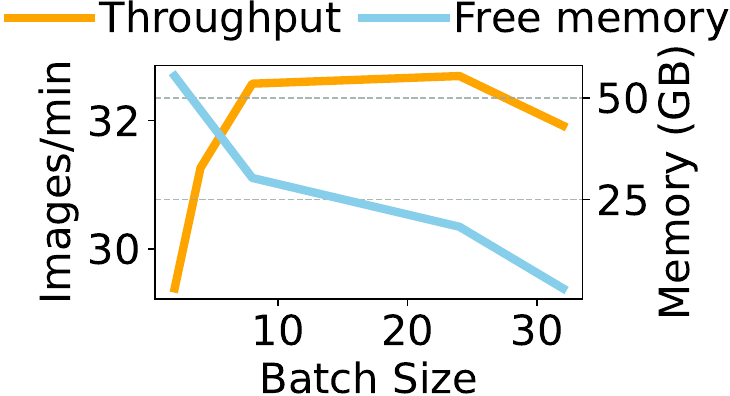}
        \caption{\small Stable-diffusion-3}
    \label{fig:sd3_memory}
    \end{subfigure}
    \begin{subfigure}[b]{0.22\textwidth}
        \includegraphics[width=\columnwidth]{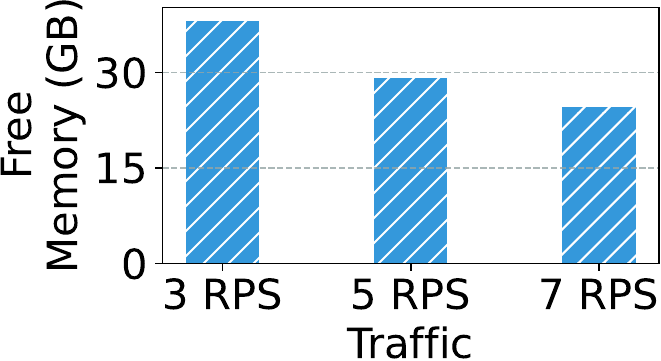}
        \caption{\small Llama 3.1 8B}
    \label{fig:llama_memory}
    \end{subfigure}
    \caption{\small Figure~\ref{fig:sd3_memory} shows the profile generated by \sysname on Stable-diffusion-3 which reaches peak throughput around a batch size of 8 Images with close to 30 GB of unused memory on the GPU. Figure~\ref{fig:llama_memory} shows that at a steady rate of 5 RPS which adheres to the SLA, there is more than 25 GB of free memory on Llama 8B.}
    \label{fig:profile_memory}
\end{figure}

\myparab{How are models classified?} 
Figure~\ref{fig:sd3_memory} presents the trace generated by \sysnameprofiler for the Stable-diffusion-3 model on an H100 GPU. The profiler increases the batch size during inference until throughput saturates, recording the corresponding free GPU memory. Figure~\ref{fig:llama_memory} shows the trace for the Llama 3.1 8B model. Following prior work~\cite{sarathi_serve,vllm}, we select an SLA with a TTFT under 1 second and an inter-token latency (ITL) below 100 milliseconds, targeting interactive workloads with a batch size of 512. Based on this SLA and ShareGPT input request distribution, our profiler tests various request rates and captures the available free GPU memory during serving. Our experiments found that the maximum sustainable traffic for the LLM within the SLA was 5 requests per second (5 RPS) shown in Figure~\ref{fig:profile_memory}. We similarly profile all other models in our experimental setup to classify producers and consumers. 


\myparab{Placing models on GPUs.} 
After profiling, we sample 10\% from audio models, 15\% from vision models and the remainder from LLMs (to represent the higher popularity of LLMs) from the models in Table~\ref{tab:models}, till we exhaust the available GPUs. We execute \sysnamePl with the sampled models and their memory profiles from \sysnameprofiler.

\begin{figure}[h]
    \centering
    \begin{subfigure}[b]{0.22\textwidth}
        \includegraphics[width=\columnwidth]{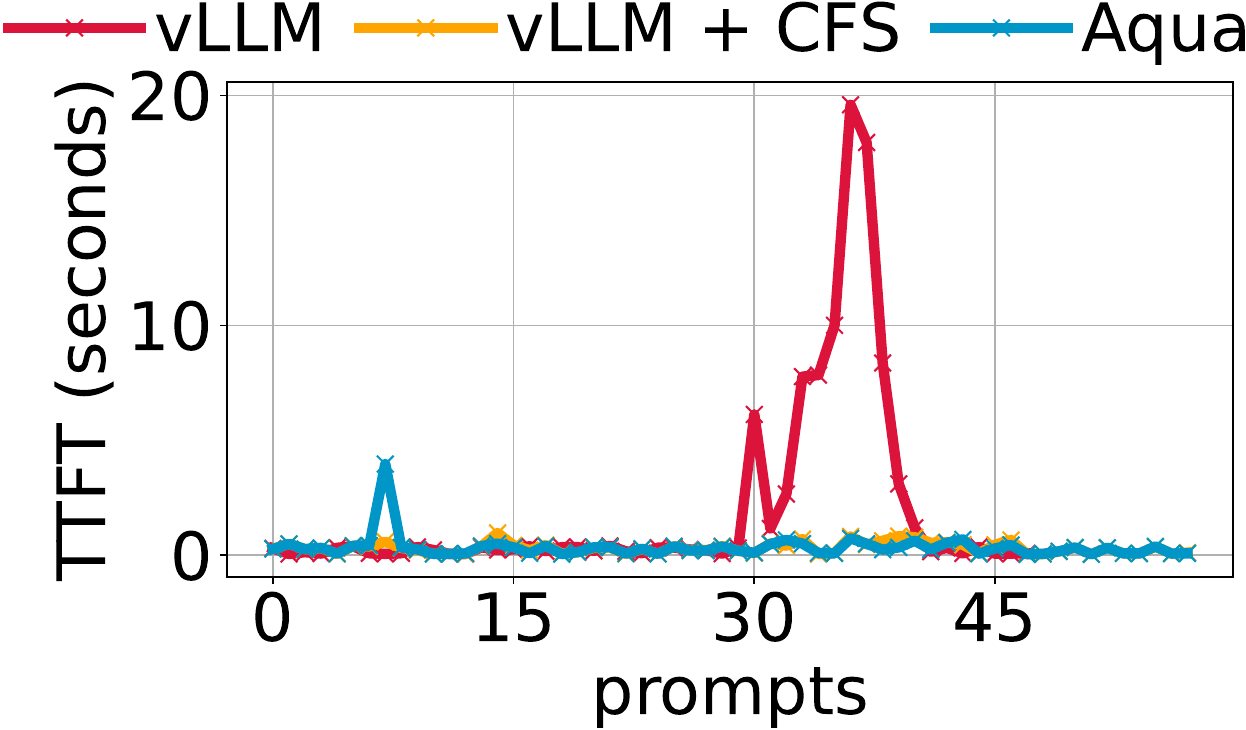}
        \caption{\small TTFT}
    \label{fig:cfs_ttft}
    \end{subfigure}
    \begin{subfigure}[b]{0.22\textwidth}
        \includegraphics[width=\columnwidth]{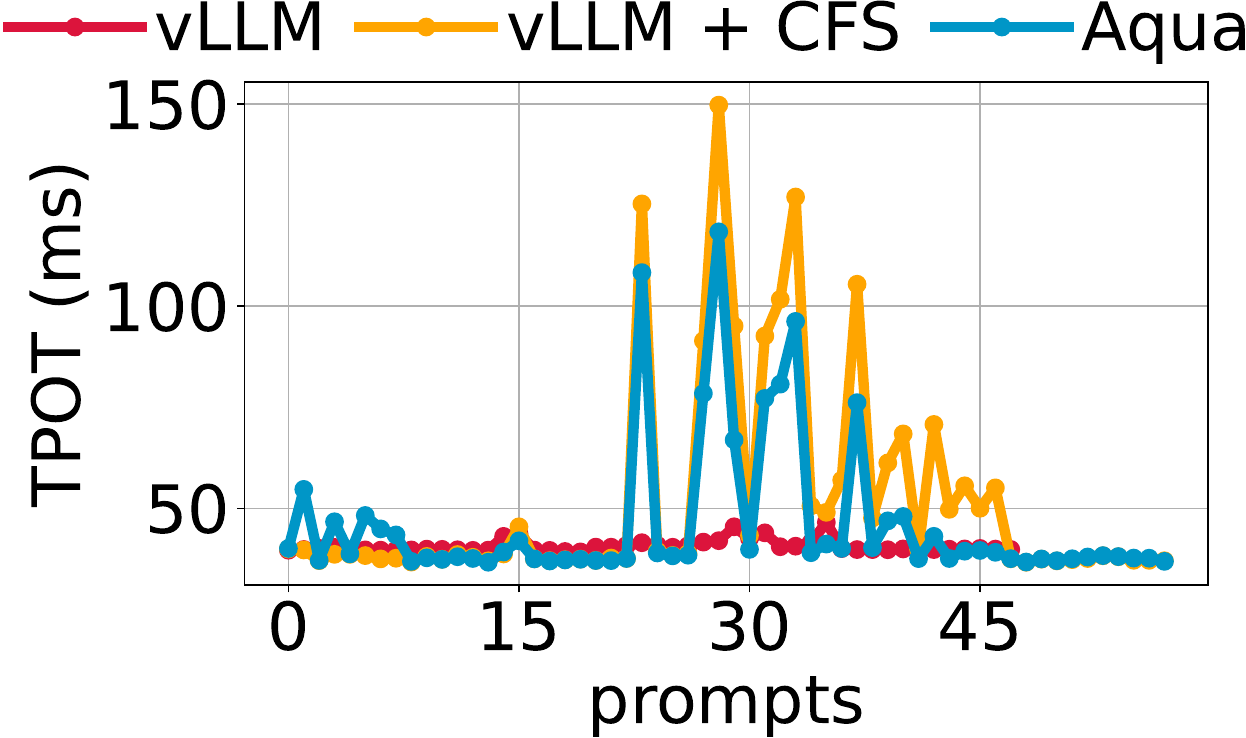}
        \caption{\small TPOT}
    \label{fig:cfs_topo}
    \end{subfigure}
    \caption{\small Figure~\ref{fig:cfs_ttft} shows the TTFT of the requests arriving in order to Llama 3.1 70B running on 2 GPUs using tensor parallelism. The requests are sampled from ShareGPT. Figure~\ref{fig:cfs_topo} shows the TPOT of the requests in order. \sysname minimizes the TTFT while incurring up to $1.5\times$ lower TPOT compared to vLLM + CFS.}
    \label{fig:cfs_benefits}
\end{figure}

\myparab{How much does CFS benefit from \sysname?}
\sysnamePl matched the distributed CFS workload running Llama 3.1 70B on 2 GPUs with two single-GPU models, Llama 8B and Audiogen. The CFS workload's two shards use \sysnamelib to allocate swap space, transparently placed on the GPUs running Llama 8B and Audiogen. We start the workload with stable request rate for CFS and after issuing 25 prompts at this rate, the workload generator doubles the request rate for a minute. This surge, handled by the baseline FCFS scheduler, results in high TTFT of 20 seconds, as shown in Figure~\ref{fig:cfs_ttft}.

A CFS policy with vLLM reduces TTFT but incurs a high TPOT from preemption overheads to DRAM (Figure~\ref{fig:cfs_topo}). In contrast, \sysname not only reduces TTFT by $20\times$ but also reduces TPOT by up to $1.5\times$ compared to vLLM + CFS. \sysname shortens the burst impact as well: in Figure~\ref{fig:cfs_ttft}, TPOT remains high between 25-48 requests with vLLM + CFS, but only between 25-40 with \sysname. This improvement stems from \sysnamelib's fast preemption, which transparently offloads data via \nvlinks to neighboring GPUs. Figure~\ref{fig:cfs_benefits} also shows that the behavior of CFS is identical to FCFS when there is no memory contention, during the first 25 and after 50 requests.

\begin{figure}[h]
    \centering
    \begin{subfigure}[b]{0.27\textwidth}
        \includegraphics[width=\columnwidth]{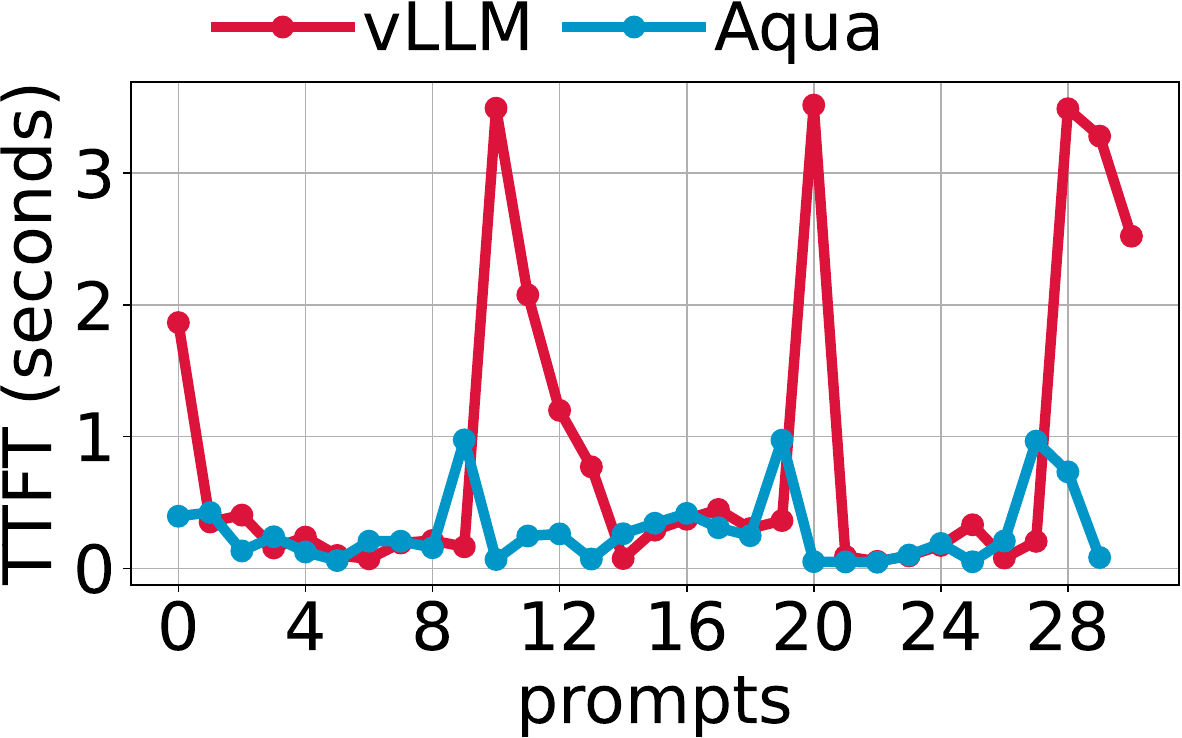}
        \caption{\small TTFT on Yi-34B-200K}
    \label{fig:prefill_ttft}
    \end{subfigure}
    \begin{subfigure}[b]{0.17\textwidth}
        \includegraphics[width=\columnwidth]{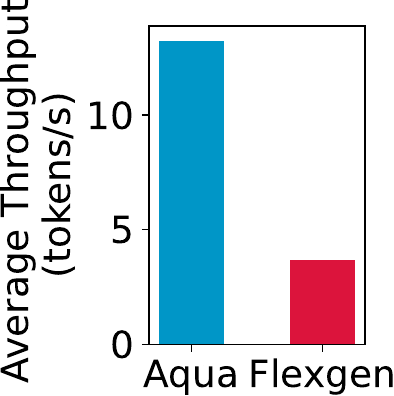}
        \caption{\small OPT-30B}
    \label{fig:opt_throughput}
    \end{subfigure}
    \caption{\small Figure~\ref{fig:prefill_ttft} shows the TTFT of requests in order sent to Yi-34B-200K served on 2 GPUs with tensor parallelism, where spikes occur due to loading cached context from swap space. Figure~\ref{fig:opt_throughput} shows the decoding throughput of OPT-30B serving on 1 GPU. \sysname significantly improves performance of both workloads.}
    \label{fig:prefill_opt}
\end{figure}

\myparab{Improvements on prefill caching.}
\sysnamePl matched the distributed Yi-34B-200K running on 2 GPUs with two single-GPU producers, Stable-Diffusion-3 (SD-3) and Kandinsky (Kand). We start with a stable request rate and issue a 100K token prompt, using an API to cache the inference context for all but the last 256 tokens. Then, 10\% of requests use the cached tokens as a prefix with different 256-token suffixes, while the rest are sampled from ShareGPT. The serving engine computes the context initially, taking around 10 seconds, caches it in the swap space, and loads it into the GPU when a request with the same prefix arrives. Figure~\ref{fig:prefill_ttft} traces 30 steady-state requests after caching.

The baseline sees TTFT spikes up to 4 seconds on each cache hit, with subsequent requests also facing high TTFT as they are batched with the long prompt and wait for loading its context into the GPU. \sysnamelib minimizes the context loading time from swap space to GPU, reducing the long prompt's TTFT by $3.5\times$. This also lowers the TTFT for subsequent requests batched with the cached prompt.

\myparab{Improvements on memory constrained inference.}
\sysnamePl matched OPT-30B running on a single GPU with Llama 3.1 8B. We run OPT-30B with FlexGen which is used for offline inference scenarios~\cite{openai_batch} with low resources and aims to improve the throughput of the system by efficiently swapping context to the DRAM. We measure the generation throughput of FlexGen on a sequence of prompts of length 8192 for 10 minutes and report the average throughput across prompts. The context generated by the prompt with 8K tokens is too large to fit on the GPU after loading the model weights and the system starts swapping to execute inference. Figure~\ref{fig:opt_throughput} demonstrates that \sysname consistently achieves a $4\times$ improvement in generation throughput compared to baseline swapping to DRAM throughout the experiment.

\begin{figure}[h]
    \centering
    \begin{subfigure}[b]{0.22\textwidth}
        \includegraphics[width=\columnwidth]{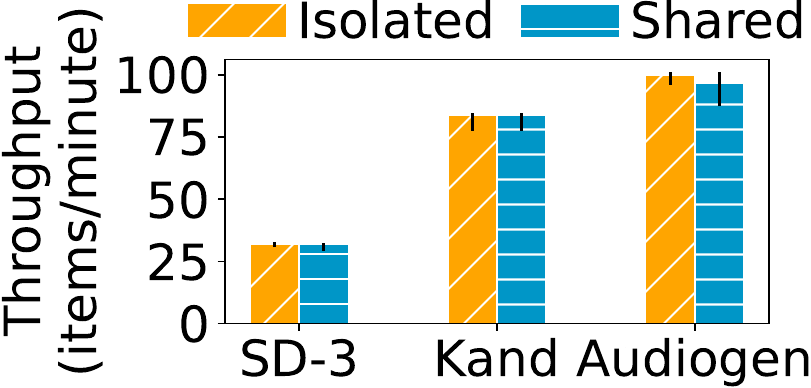}
        \caption{\small Throughput}
    \label{fig:throughput_impact}
    \end{subfigure}
    \begin{subfigure}[b]{0.22\textwidth}
        \includegraphics[width=\columnwidth]{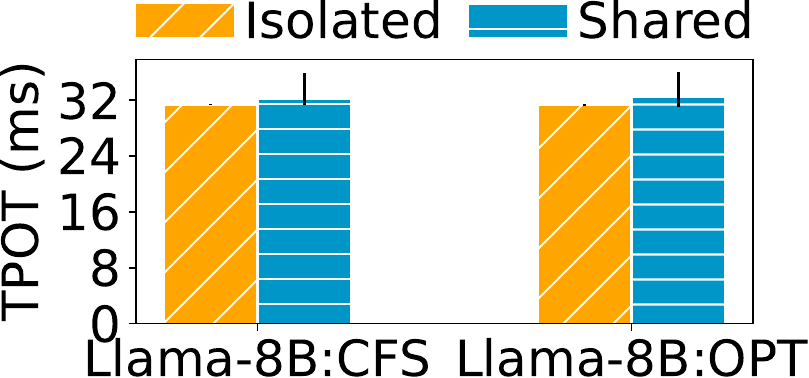}
        \caption{\small TPOT}
    \label{fig:tpot_impact}
    \end{subfigure}
    \caption{\small Figure~\ref{fig:throughput_impact} shows negligible impact on throughput when audio and image producers share memory with consumer workloads. The isolated bars reflect producer performance without consumers. Figure~\ref{fig:tpot_impact} shows that memory sharing increases the average TPOT of 2 Llama 8B instances by about 4.5 milliseconds.}
    \label{fig:impact}
\end{figure}

\myparab{What is the impact of sharing memory on producers?}
We executed all the producers that shared memory in the previous experiments without consumers to compare it with the performance when sharing memory with a consumer. Figure~\ref{fig:impact} shows that the impact of sharing memory is low for producers of all modalities. The difference in throughput is less than 5\% for audiogen and non-existent for vision generative models. We use average TPOT to measure the performance of LLM producers since it captures the time taken to execute one inference iteration. Figure~\ref{fig:tpot_impact} shows that the impact of sharing memory is about 4.5 milliseconds on TPOT. A consumer accessing the producer's GPU memory could lead to memory bandwidth contention, as shown in recent work~\cite{hostcongestion}. We posit that the reason for the reduction in performance of LLMs and audiogen models is because they rely on transformer layers that are known to be bottlenecked by memory bandwidth of the GPU~\cite{sarathi_serve} and any bandwidth contention will reduce the performance. 

\begin{figure}[h]
    \centering
    \begin{subfigure}[b]{0.18\textwidth}
        \includegraphics[width=\columnwidth]{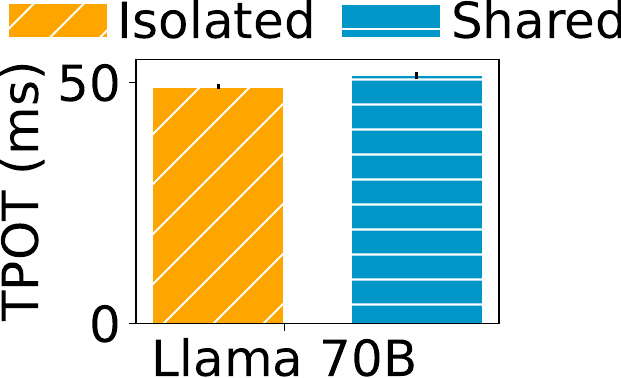}
        \caption{\small Llama-70B TPOT}
    \label{fig:dist_tpot_impact}
    \end{subfigure}
    \begin{subfigure}[b]{0.25\textwidth}
        \includegraphics[width=\columnwidth]{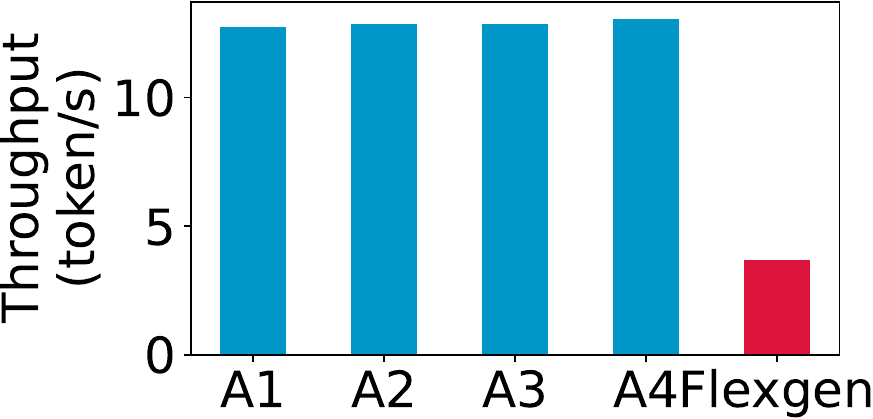}
        \caption{\small OPT-30B throughput}
    \label{fig:dist_throughput}
    \end{subfigure}
    \caption{\small Figure~\ref{fig:dist_tpot_impact} shows that the impact of sharing memory on the average TPOT over the duration of 10 minutes is 5\% on Llama 70B served on 4 GPUs using tensor parallelism on the ShareGPT dataset. Figure~\ref{fig:dist_throughput} shows the throughput of the 4 OPT-30B instances during the same duration which is consistently better than the baseline by $4\times$. (A=\sysname)}
    \label{fig:dist_impact}
\end{figure}

\myparab{What is the impact on distributed producers?}
Since \sysnamePl did not match any consumer with a distributed LLM producer, we study the impact of sharing memory over \nvlinks on a distributed LLM's performance in this experiment. Inference with tensor parallelism requires inter-GPU communication via \nvlinks, which are also used when consumers access memory on the producer's GPUs, potentially affecting performance. To assess this, we run the Llama 3.1 70B model across 4 GPUs with tensor parallelism and 4 instances of OPT-30B on another 4 GPUs. Each GPU serving Llama has 15 GB of free space and is a producer to one of the OPT-30B which needs 10 GB swap space to infer on an 8K tokens prompt. We pick OPT-30B for this experiment since it frequently utilizes the \nvlinks to swap inference context.

We ran the experiment for 10 minutes, with Llama handling steady requests and OPT-30B generating tokens from long prompts. Figure~\ref{fig:dist_impact} shows that sharing memory results in a negligible TPOT overhead on Llama is 5\%. We investigated the cause for the low overhead and determined two reasons: first, the communication size during tensor parallelism is only a few megabytes. Second, this communication occurs between layers, leaving the interconnect idle during computation. The small buffer sizes combined with fact that the interconnect is idle during computation, leads to the underutilization of \nvlinks. As a result, consumer's traffic barely result in any additional impact. We believe that this overhead can be eliminated if NVIDIA exposes an API to set priority on the packets which we can use to prioritize tensor parallelism packets over producer-consumer traffic.

\subsection{\sysname's adaptability}
\label{subsec:adaptability}
In this subsection, we demonstrate \sysname's adaptability to varying producer workload demands and \sysname's extensibility to different datasets, using the A100 testbed.

\begin{figure}[h]
    \centering
    \begin{subfigure}[b]{0.22\textwidth}
    \includegraphics[width=\columnwidth]{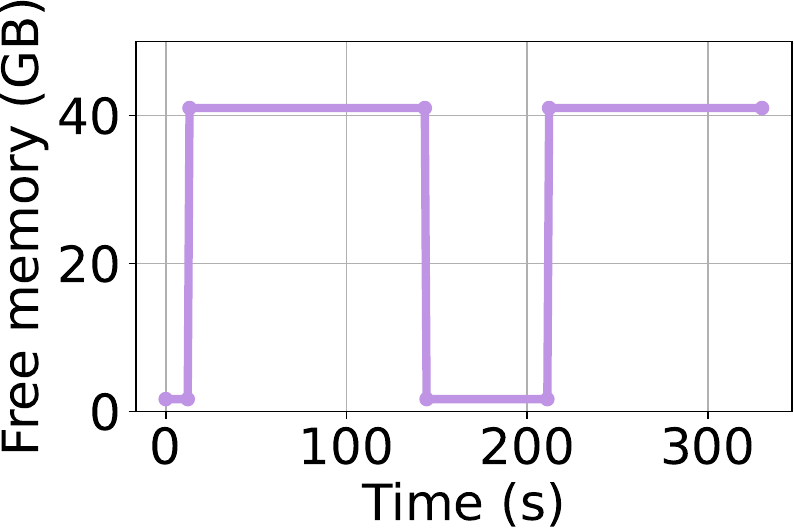}
    \caption{\small Free memory (Llama 8B)}
    \label{fig:elastic_memory}
    \end{subfigure}
    \begin{subfigure}[b]{0.22\textwidth}
    \centering
    \includegraphics[width=\columnwidth]{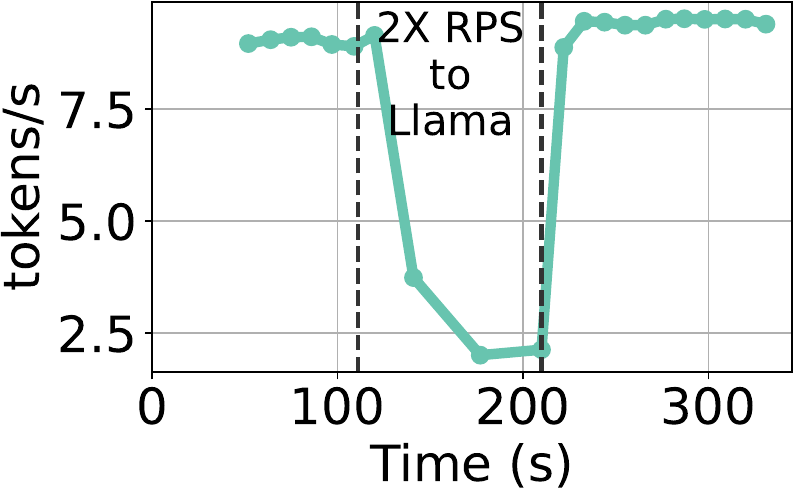}
    \caption{\small OPT-30B throughput.}
    \label{fig:elastic_throughput}
    \end{subfigure}
    \caption{\small \sysname's elasticity opportunistically improves the throughput of a memory constrained workload by $6\times$.}
    \label{fig:elastic_result}
\end{figure}

\myparab{\sysname's elasticity.}
In this experiment, we host Llama 3.1 8B as 
producer and OPT-30B as consumer 
job with \sysname on a GPU server with 2 A100 GPUs. 
We vary the number of inference requests sent 
to the producer. Figure~\ref{fig:elastic_memory} shows 
the available free memory on the GPU serving Llama 8B. At the start,
the serving engine reserves all the memory to
accommodate the inference context of incoming requests. 
\sysnamelib notices that the inference load is steady and 
using the profiled data, it offers memory on this GPU to consumers.

Near the 60-second mark in Figure~\ref{fig:elastic_memory}, we begin 
the long prompt inference, achieving a high throughput of 8 tokens per second. 
At the 120 seconds mark, we double the traffic to the producer. 
This increases the number of requests in the 
pending queue and \sysnamelib notices this and 
rapidly reclaims memory to accommodate the load. 
Reclaiming memory reduces
throughput of the consumer (Figure~\ref{fig:elastic_throughput}). 
The throughput again increases when LLM is 
done serving all the requests.
\sysnamelib again detects the steady traffic and 
transparently moves the offloaded tensors of the consumer 
back to the producer's GPU. This shows that \sysname can
dynamically respond to changes in producer 
workloads and opportunistically improve consumer performance.

\begin{figure}[h]
    \centering
    \begin{subfigure}[b]{0.22\textwidth}
        \includegraphics[width=\columnwidth]{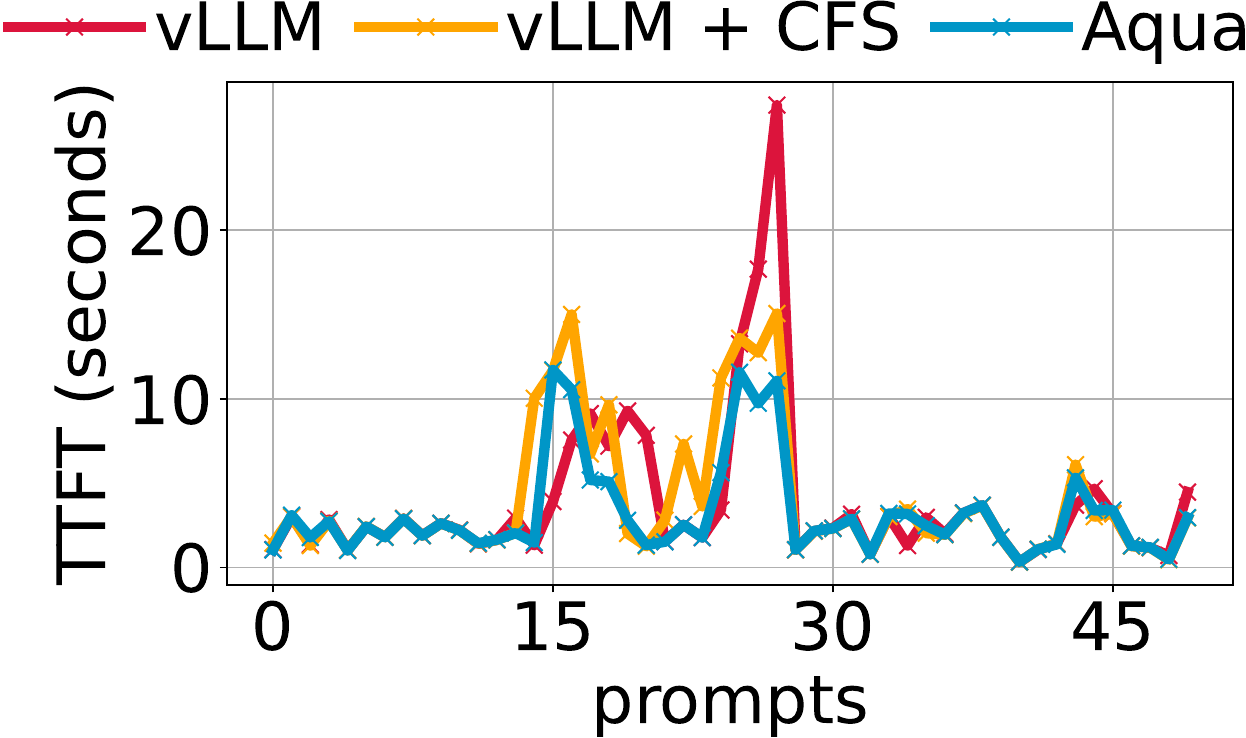}
        \caption{\small TTFT}
    \label{fig:arxiv_cfs_ttft}
    \end{subfigure}
    \begin{subfigure}[b]{0.22\textwidth}
        \includegraphics[width=\columnwidth]{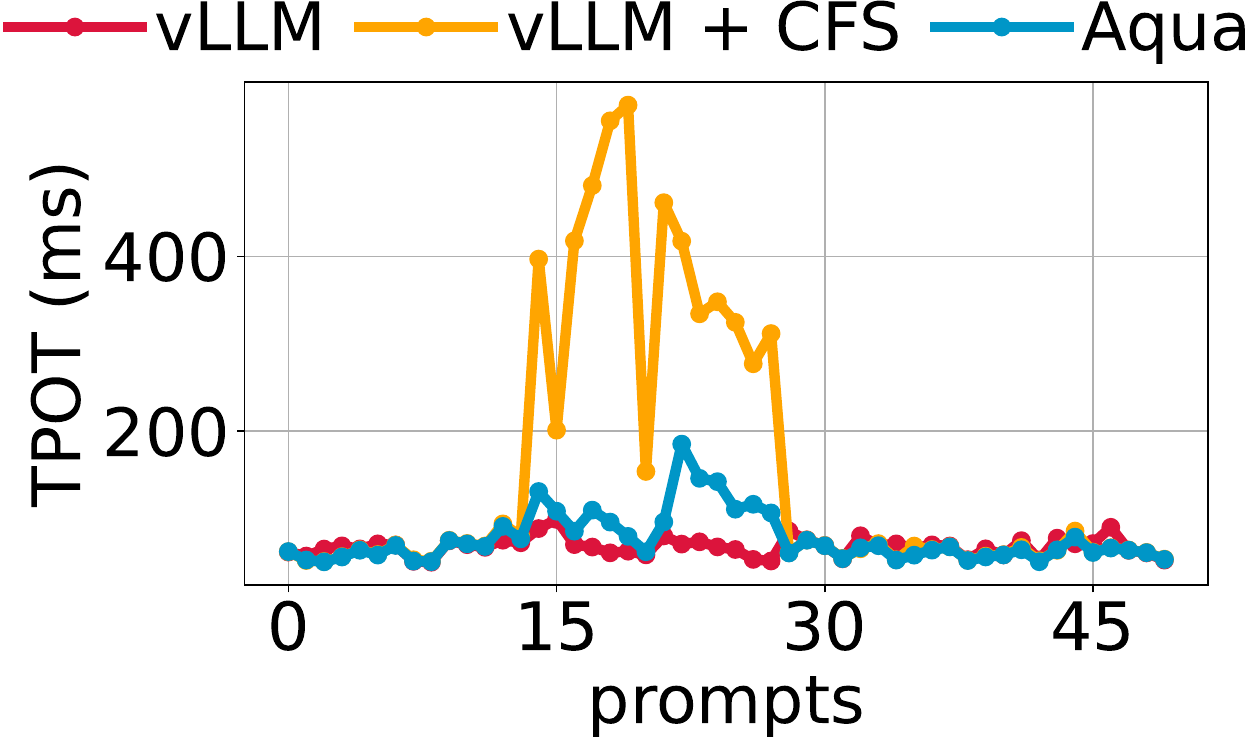}
        \caption{\small TPOT}
    \label{fig:arxiv_cfs_topo}
    \end{subfigure}
    \caption{\small Figure~\ref{fig:arxiv_cfs_ttft} shows the TTFT of the requests arriving in order to Yi-34B running on 1 A100 GPU. The requests are sampled from Arxiv dataset. Figure~\ref{fig:arxiv_cfs_topo} shows the TPOT of the requests.}
    \label{fig:arxiv_cfs_benefits}
\end{figure}

\myparab{\sysname's benefits extend to different datasets.}
In this experiment, we use the arxiv summarization dataset and Yi-34B to show how \sysname's
benefits carry over to other datasets.
The arxiv dataset has longer prompts, a median prompt is around 7000 tokens compared to 
the 2000 tokens in ShareGPT dataset. So, we start by using \sysnameprofiler to profile
the swap space requirement for Yi-34B executing on a single A100 GPU to handle bursts.
Given the stable request rate, the profiler creates bursts that are multiples of 
the stable request rate and determines that the model requires 6GB, 15GB 
and 40GB of swap space to handle bursts for a minute that are $2\times$, $3\times$ and $5\times$ the stable request rate.  

We begin by hosting Llama 8B on one GPU and Yi-34B on another, issuing requests at a stable rate. To simulate a burst, we triple the request rate to Yi-34B for one minute. Figure~\ref{fig:arxiv_cfs_benefits} shows that \sysname efficiently handles the burst, reducing peak TTFT by $2\times$ and lowering TPOT by $2\times$ through fast preemption. TTFTs for prompts between 10-20 are higher across all systems because prompts are long and the prefill stage is mainly bottlenecked by compute.

\section{Related work}

\myparab{GPU job schedulers.} 
Usher~\cite{usher} is the most recent scheduler that profiles compute and memory utilization on a GPU and multiplexes complementary models on the GPU to increase resource utilization. Alpa-serve~\cite{alpa_serve} dynamically scales the GPUs allocated to the job by allocating underutilized GPUs of other workloads to jobs with high traffic rates. Similarly, Salus \cite{salus}, clockwork \cite{clockwork}, Orion \cite{orion} and PipeSwitch \cite{pipeswitch} multiplex models at the model, layer and kernel granularity to efficiently time share the compute on the GPU. All these systems tightly couple compute and memory allocation on a GPU while multiplexing models and miss the opportunity to utilize the free memory capacity when the compute is saturated. \sysname bridges this gap and maximizes the memory utilization on multi-GPU servers even when there is no opportunity to share compute.

\myparab{Schedulers for online LLM inference.} 
Recently, the community has proposed many algorithms for scheduling prompts arriving at an LLM~\cite{sarathi_serve,vtc,llumnix,mlfq}. Sarathi-serve~\cite{sarathi_serve} proposed a scheduler to batch prefills and decodes. Llumnix~\cite{llumnix} schedules prompts across LLM instances by load balancing, auto-scaling and live-migrating prompts. VTC~\cite{vtc} does coarse grained fair scheduling for LLM serving engines via admission control to prevent bad actors from overwhelming the infrastructure. VTC also mentions that preemptive scheduling is orthogonal to their work and should be addressed in the future. CFS in \sysname complements these systems by handling bursts while auto-scaling kicks in and by fairly allocating resources to admitted prompts. Efforts like MLFQ~\cite{mlfq} also build preemptive schedulers and \sysnamelib can improve their performance even more with fast preemption. 

\myparab{CUDA Unified virtual memory.} 
{
    CUDA unified memory provides a single virtual address space spanning GPUs and the host memory. The CUDA driver manages page faults and prefetching allowing GPUs to allocate and access memory seamlessly across all the GPUs and DRAM. However, its default prefetching strategy is insufficient for LLM inference, where inference contexts are scattered across non-contiguous memory regions. Moreover, the CUDA driver does not opportunistically utilize idle GPU memory across the NVLink and NVSwitch interconnect. Prior work, such as Memory Harvester~\cite{memory_harvesting_nvlinks}, addresses this limitation by leveraging idle GPU memory as swap space at the driver level but remains application-agnostic, giving workloads no control over eviction and prefetching. This lack of control can lead to inefficient paging, such as evicting model weights instead of inference context, increasing overhead. In contrast, LLM serving engines manually manage prefetching and eviction at the application layer to optimize memory usage. \sysname improves upon CUDA unified memory by enabling explicit control over prefetch order and opportunistically migrating offloaded data to faster GPU memory, maximizing paging performance for LLM inference.
}

\myparab{Other generative engines.} 
Orca \cite{orca} is an LLM inference engine that introduced continuous batching, 
while vLLM \cite{vllm} improved this by designing paged attention. 
Deepspeed-zero \cite{deepspeed_zero} was the baseline for
FlexGen \cite{flexgen}. Given that 
\sysname can enhance FlexGen's performance, similar 
benefits can be extend to Deepspeed. HeteGen \cite{hetegen2024} 
builds upon FlexGen by utilizing both the CPU and 
GPU for computation. Since HeteGen also relies 
on PCIe for CPU to GPU I/O, \sysname is complementary to 
it and can further improve its performance. 
Unlike CacheGen~\cite{cache_gen}, which trades accuracy 
for compression and stores many 
prompts in remote memory, \sysname caches contexts while 
maintaining accuracy which is suitable for scenarios that use 
critical data like healthcare, legal and finance. 
We note that there are simulators~\cite{vidhur} facilitating fast profiling. \sysnameprofiler can also interface with such simulators and profile even faster but we measure directly from the hardware in this work for accuracy. Dist-serve~\cite{dist_serve} and Splitwise~\cite{splitwise} propose using different hardware for the prefill and decoding stage. As \sysnameprofiler can classify disaggregated workloads as producers or consumers, \sysname is complementary to them.

\myparab{Swapping DRAM pages on RDMA.} 
Infiniswap \cite{infiniswap} and Fastswap \cite{fastswap}
accelerate paging in Operating Systems by harvesting memory and moving pages across DRAM of interconnected servers over RDMA. The memory allocation algorithms in these systems do not guarantee bandwidth isolation since 
the same producer can be shared across multiple consumers. Understanding the behavior of closed systems like \nvlinks and NVSwitch under contention is critical to provide paging bandwidth guarantees but is challenging since their congestion control algorithm is closed-source. Even though \sysname's design allows us to implement similar memory allocation algorithms~\cite{infiniswap,fastswap} to allocate memory across GPUs, we developed \sysnamePl because we want to provide bandwidth guarantees during paging. We leave the exploration of modeling paging bandwidth over \nvlinks during contention to future work. 

\section{Acknowledgements}
This work was supported in part by ACE, one of the seven centers in JUMP 2.0, a Semiconductor Research Corporation (SRC) program sponsored by DARPA.
The authors of this work are also supported by NSF Award \#2444537 and the Cornell Bowers CIS-LinkedIn Grant.

\bibliographystyle{ACM-Reference-Format}
\balance
\bibliography{references}

\appendix
\section{Artifact Appendix}

\subsection{Abstract}

We are releasing the source code of \sysname at \url{https://github.com/aquaml}. \sysname needs a multi-GPU server with at least 2 NVIDIA GPUs interconnected with \nvlinks. \sysname's documentation is available at \url{https://aquaml.github.io/}.

\subsection{Artifact check-list (meta-information)}

{\em Here is a list of the repositories that are available in the artifact.}

{\small
\begin{itemize}
  \item {\bf \sysnamelib which includes \sysnameT.}
  \item {\bf vLLM with \sysname's CFS algorithm}
  \item {\bf \sysnamePl.}
  \item {\bf PyTorch scripts with integrated \sysnamelib to execute stable-diffusion.}
  \item {\bf FlexGen with integrated \sysnamelib.}
  \item {\bf Audiogen with integrated \sysnamelib.}
  \item {\bf Publicly available?: Yes}
\end{itemize}
}

\end{document}